\documentclass[a4paper,twocolumn,nofootinbib]{revtex4-1}

\usepackage{xcolor}
\usepackage{hyperref}
\hypersetup{
	colorlinks,
	linkcolor={red!90!black},
	citecolor={black!10!blue},
	urlcolor={blue!80!black}
}

\usepackage{tikz}
\usepackage{amssymb}
\usepackage{bm,amsmath}
\usepackage{graphicx}
\usepackage{array}

\newcolumntype{C}[1]{>{\centering\arraybackslash}p{#1}}

\newcommand{\bx}{\mathbf{x}}
\newcommand{\mK}{\mathbf{K}}

\newcommand{\mL}{\mathbf{\Lambda}}

\newcommand{\g}{\gamma}
\renewcommand{\L}{\Lambda}
\newcommand{\rav}{\bar{r}}

\begin{document}
	
\title{The origin of universal cell shape variability in a confluent epithelial monolayer}
\author{Souvik Sadhukhan}
\email{ssadhukhan@tifrh.res.in}
\affiliation{TIFR Centre for Interdisciplinary Sciences, Tata Institute of Fundamental Research, Hyderabad - 500046, India}

\author{Saroj Kumar Nandi}
\email{saroj@tifrh.res.in}
\affiliation{TIFR Centre for Interdisciplinary Sciences, Tata Institute of Fundamental Research, Hyderabad - 500046, India}




\begin{abstract}

Cell shape is fundamental in biology. The average cell shape can influence crucial biological functions, such as cell fate and division orientation. But cell-to-cell shape variability is often regarded as noise. In contrast, recent works reveal that shape variability in diverse epithelial monolayers follows a nearly universal distribution. However, the origin and implications of this universality are unclear. Here, assuming contractility and adhesion are crucial for cell shape, characterized via aspect ratio (AR), we develop a mean-field analytical theory for shape variability. We find that a single parameter, $\alpha$, containing all the system-specific details, describes the probability distribution function (PDF) of AR; this leads to a universal relation between the standard deviation and the average of AR. The PDF for the scaled AR is not strictly but almost universal. The functional form is not related to jamming, contrary to common beliefs, but a consequence of a mathematical property. In addition, we obtain the scaled area distribution, described by the parameter $\mu$. We show that $\alpha$ and $\mu$ together can distinguish the effects of changing physical conditions, such as maturation, on different system properties. The theory is verified in simulations of two distinct models of epithelial monolayers and agrees well with existing experiments. We demonstrate that in a confluent monolayer, average shape determines both the shape variability and dynamics. Our results imply the cell shape variability is inevitable, where a single parameter describes both statics and dynamics and provides a framework to analyze and compare diverse epithelial systems. 

\end{abstract}

\maketitle

D'Arcy Thompson argued, in his book {\it On Growth and Form}, physical principles could explain tissue packing and cell shape \cite{d'arcybook}. Shape formation of tissues and organs during embryogenesis is a long-standing, fascinating problem of developmental biology. Since cells are the functional units of a tissue, shapes in the organs must originate at the cellular level \cite{paluch2009,wyatt2015,gonzalez2021,hannezo2013}. Cell shapes are vital in both health and disease. As cancer progresses \cite{sailem2017,park2016}, as asthma advances \cite{park2015,park2016,veerati2020,atia2018}, as wounds heal \cite{nnetu2012,poujade2007}, as an embryo develops \cite{farhadifar2007,atia2018}, cells progressively change their shape. Besides, cell shape may influence crucial biological functions, such as cell growth or selective programmed cell death (apoptosis) \cite{chen1997}, the orientation of the mitotic plane \cite{wyatt2015,bosveld2016,xiong2014}, stem cell lineage \cite{mcbeath2004,wang2011}, terminal differentiation \cite{watt1988,roskelley1994}, and division-coupled interspersion in many mammalian epithelia \cite{mckinley2018}. Moreover, the nuclear positioning mechanism in neuroepithelia depends on cell shape variation \cite{yanakieva2019}. Thompson regarded cell-to-cell shape variability as a biologically unimportant noise \cite{d'arcybook}; however, it is now known that shape variability is not an exception but a fundamental property of a confluent cellular monolayer \cite{graner2017}. In a seminal work, Atia {\it et al} \cite{atia2018} showed that cell shape variability, quantified by the aspect ratio (AR), follows virtually the same distribution across different epithelial systems. But, the origin of this near-universal behavior, and whether it is precisely universal, remains unclear.

Reference \cite{atia2018} argued that this behavior in an epithelial monolayer is related to the jamming transition. Previous works have established the similarity in dynamics between cellular monolayers and glassy systems \cite{angelini2011,park2015,malinverno2017}. The glassy dynamics in confluent systems has been argued, via vertex model (VM) simulations, to be controlled by a rigidity transition akin to the jamming transition \cite{bi2015,park2015}. In a jammed granular system, the statistics of tessellated volume, $x$, follows a $k$-Gamma distribution, $P(x,k)$, where $k$ is a parameter. From this association, Ref. \cite{atia2018} conjectured the viability of describing the AR distribution via $P(x,k)$. However, such an association has several problems: (1) The glassy dynamics in a cellular monolayer is also reasonably described by Voronoi and cellular Potts models (CPM) \cite{bi2016,souvik2021}. In these models, in contrast to the VM, there is no rigidity transition \cite{sussmansoftmatter2018,souvik2021}. Hence, the relevance of jamming physics to an epithelial system remains unclear. (2) More important, the experiments and simulations in Ref. \cite{atia2018} show that this distribution persists even in the fluid regime, where jamming physics is not applicable even within the VM. (3) The analytical derivation of the $k$-Gamma distribution in granular packing \cite{aste2008} relies on the fact that $x$ is additive, whereas, as the authors of Ref. \cite{atia2018} rightly point out, AR, $r$, is not. Thus, there {\em exists no rigorous basis for the applicability of the $k$-Gamma distribution} \cite{atia2018}.

Yet, Ref. \cite{atia2018} and several subsequent works \cite{lin2018,li2020,kim2020,wenzel2021} have shown that the probability distribution function (PDF) of scaled AR, $r_s$, in diverse biological and model systems is described by $P(r_s,k)$, and the value of $k$ is almost universal, around $2.5$. Furthermore, the standard deviation, $sd$, vs the mean AR, $\rav$, follows a universal relation \cite{atia2018,kim2020,ilina2020}. What is the origin of this universality? What determines the value of $k\sim 2.5$? How is the PDF of $r$ related to the microscopic properties of a system? Answers to questions like these are crucial for deeper insights into the cell shape variability and unveiling the implications of the universality. However, it requires an analytical theory that is rare in this field due to the inherent complexity of the problem and the presence of many-body interactions.

In this work, we develop a mean-field analytical theory for cell shape variability. We find that the AR distribution is described by a single parameter, $\alpha$, containing all the system-specific parameters, leading to a universal relation between $sd$ and $\rav$. On the other hand, the PDF of $r_s$ is not strictly, but virtually, universal; $k\sim 2.5$ is a direct consequence of a mathematical property. We also obtain the PDF for the scaled area, $a$, and show that it is not universal, in contrast to what has been proposed elsewhere \cite{wilk2014}. We demonstrate that simultaneous measurements of the PDFs for $a$ and $r$ can reveal the effects of changing physical conditions, such as maturation, on the individual model parameters. We have verified our theory via simulations of two distinct models of a confluent epithelial monolayer: the discrete lattice-based CPM on square and hexagonal lattices and the continuous VM (see Supplementary Material (SM), Sec. S3 for details). Moreover, comparisons with existing experimental data on a wide variety of epithelial systems show excellent agreements. We illustrate that the same parameter describes both statics and dynamics, governs the origin and aspects of universality, and, thus, provides a framework to analyze and compare diverse epithelial systems.

\subsection*{Analytical theory for the shape variability}

Simplified model systems, representing cells as polygons, have been remarkably successful in describing both the static and dynamic aspects of an epithelial monolayer \cite{honda1978,honda1980,farhadifar2007,bi2014,fletcher2014,bi2015,park2015}.
The energy function, $\mathcal{H}$, governing these models is
\begin{equation}\label{energyfunc}
\mathcal{H}=\sum_{i=1}^N\Big[\lambda_A(A_i-A_0)^2+\lambda_P (P_i-P_0)^2\Big],
\end{equation}
where $N$ is the number of cells, the first term constrains area, $A_i$, to a target area $A_0$, determined by cell height and cell volume, with strength $\lambda_A$. Experimentally, it has been shown cell height remains almost constant in an epithelial monolayer \cite{farhadifar2007}. The second term describes cortical contractility and adhesion \cite{bi2015,farhadifar2007,jacques2015}. It constrains the perimeter, $P_i$, to the target perimeter $P_0$ with strength $\lambda_P$. The energy function, Eq. (\ref{energyfunc}), can be numerically studied \cite{albert2016} via different confluent models, such as the VM \cite{farhadifar2007,bi2015}, the Voronoi model \cite{bi2016,honda1978}, or more microscopic models such as the CPM \cite{graner1992,hirashima2017,hogeweg2000} and the phase-field model \cite{nonomura2012,palmieri2015,wenzel2021}.

\begin{figure*}
	\centering
	\includegraphics[width=0.75\textwidth]{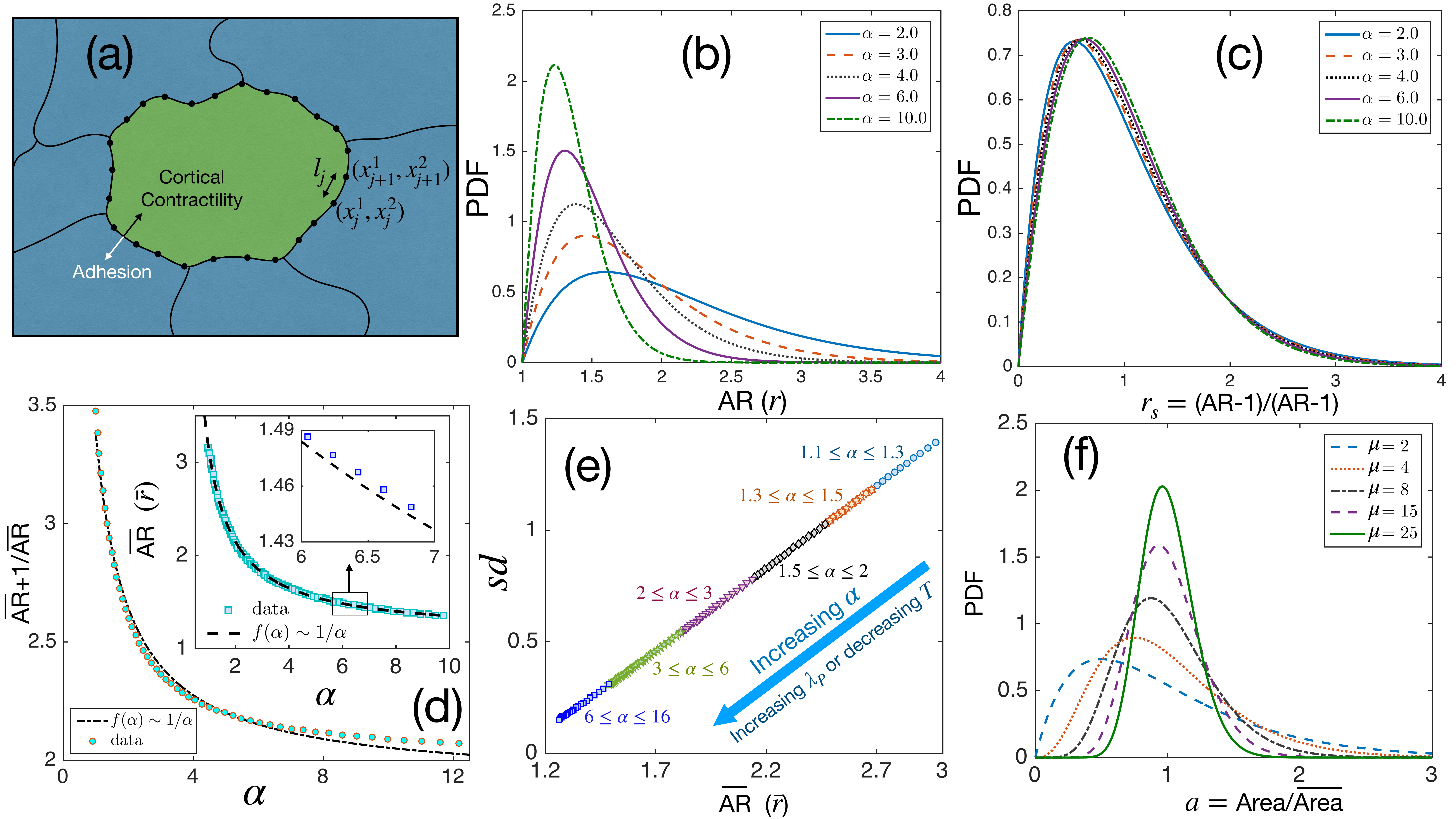}
	\caption{Theoretical results for cell shape variability. (a) Schematic illustration of the mean-field model. Cell perimeter is represented by a set of points (shown for a particular cell); $l_j$ is the distance between two neighboring points. Cortical contractility and adhesion impose competing forces. (b) PDF of aspect ratio, $r$, is governed by the parameter $\alpha$. (c) PDF of the scaled aspect ratio, $r_s=(r-1)/(\rav-1)$ is nearly universal. (d) The PDF of $r_s$ can be universal if $\rav+1/\rav$ is proportional to $1/\alpha$ (Eq. \ref{distAR}), but there is a slight deviation showing that it is not strictly universal. {\bf Inset:} $\rav$ also slightly deviates from the $1/\alpha$ form. Lines are fits with the $1/\alpha$ form. (e) Standard deviation ($sd$) vs $\rav$ follows a universal relation; the state points move towards the origin as $\alpha$ increases. (f) The PDF of $a$ follows Gamma distribution with a single parameter, $\mu$, Eq. (\ref{distarea}). }
	\label{plot_diffalpha}
	\vspace{-0.3cm}
\end{figure*}
\begin{figure*}
	\centering
	\includegraphics[width=0.9\textwidth]{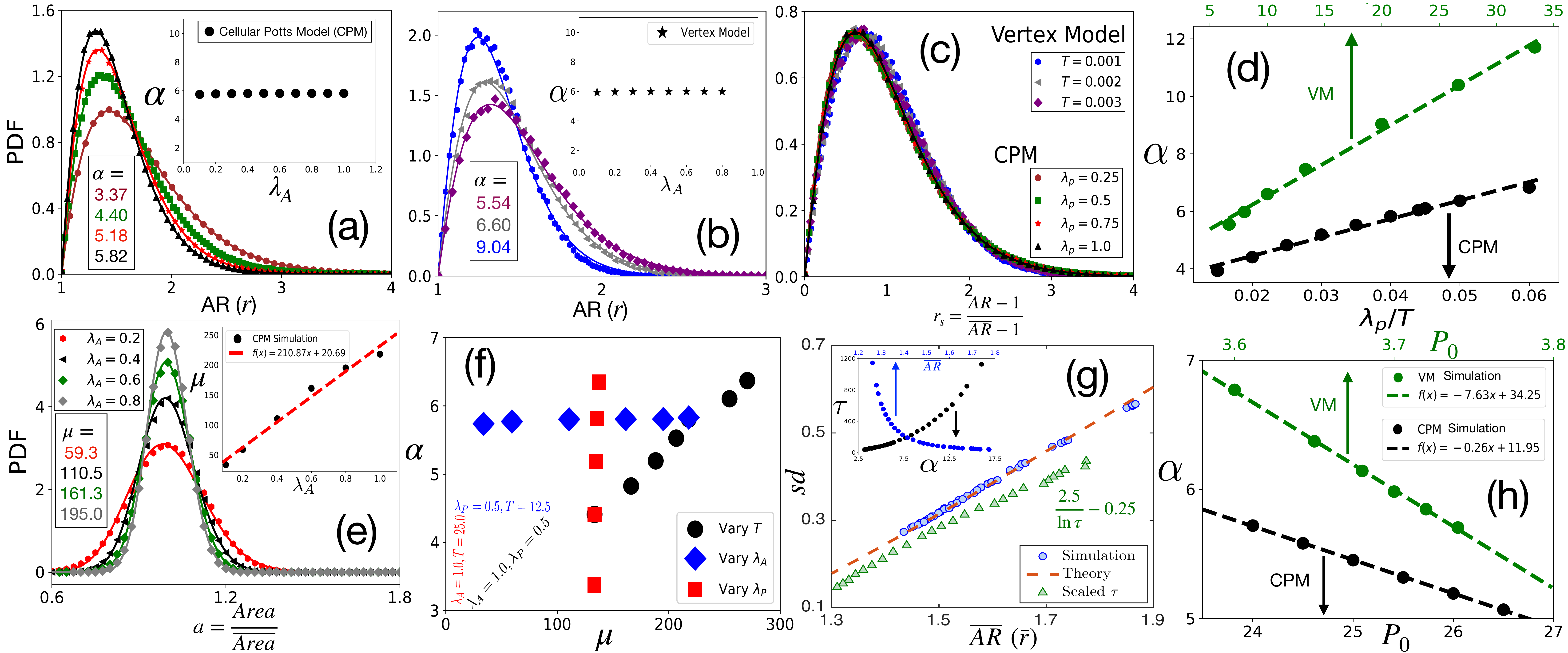}
	\caption{Comparison with simulations. (a) PDF of AR at different $\lambda_P$ (quoted in c) within the CPM, with $\lambda_A=1.0$, and $T=25.0$. Symbols are simulation data, and lines represent the fits with Eq. (\ref{distAR}). {\bf Inset:} $\alpha$ within the CPM at $\lambda_P=0.5$, and $T=25.0$ with varying $\lambda_A$ remains almost constant. (b) PDF of $r$ within the VM with varying $T$ (quoted in c), $\lambda_P=0.02$, $\lambda_A=0.5$, $P_0=3.7$. Lines are fits with Eq. (\ref{distAR}) with $\alpha$ as shown. {\bf Inset:} $\alpha$ as a function of $\lambda_A$ remains almost constant also within the VM. $T=2.5\times 10^{-3}$ and $\lambda_P=0.02$ and $P_0=3.7$. (c) PDFs for the $r_s$, for the same data as in (a) and (b) for the CPM and the VM, respectively, show a virtually universal behavior. (d) Theory predicts $\alpha$ linearly varies with $\lambda_P/T$, simulation data within both the models agree with this prediction. $\lambda_A=1.0$ for the CPM; $P_0=3.7$ and $\lambda_A=0.5$ for the VM simulations. Dotted lines are fits with a linear form. (e) PDF of the scaled area, $a$, within the CPM at different $\lambda_A$, but fixed $T=12.5$ and $\lambda_P=0.5$. The lines are fits with Eq. (\ref{distarea}) with the values of $\mu$ as shown. {\bf Inset:} $\mu$ almost linearly increases with $\lambda_A$. (f) $\alpha$ vs $\mu$ within the CPM when we have varied only one of the three variables: $T$, $\lambda_A$, and $\lambda_P$. $\alpha$ does not depend on $\lambda_A$, and $\mu$ does not depend on $\lambda_P$.
		 (g) Our theory predicts a universal behavior for $sd$ vs $\rav$, symbols are the CPM data at different parameter values, and the line is our theory (not a fit). We have also plotted the scaled relaxation time, $2.5/\ln\tau-0.25$, to show them on the same figure. $\tau$ increases as $\rav$ decreases. {\bf Inset:} $\tau$ as functions of $\alpha$ (lower axis) and $\rav$ (upper axis). (h) Theory predicts $\alpha$ linearly decreases with $P_0$, this agrees with simulations (symbols); parameters in the CPM: $\lambda_A=1.0$, $\lambda_P=0.5$, $T=16.67$; the VM: $\lambda_A =0.5$, $\lambda_P = 0.02$, $T = 0.0025$. 
[CPM simulations here are on square lattice, $\bar{A}=40$, $P_0=26$ in (a-f)]. }
	\label{simcomp}
	\vspace{-0.3cm}
\end{figure*}

We aim to develop a mean-field theory for a microscopic version, as schematically shown in Fig. \ref{plot_diffalpha}(a), to calculate the shape variability. The mechanical properties of a cell, for most practical purposes, are governed by a thin layer of cytoplasm just below the cell membrane, the cortex. Therefore, the crucial contribution should come from the perimeter term. We describe the cell perimeter via $n$ points, where $l_j$ is the infinitesimal line-element between $j$th and $(j+1)$th points (Fig. \ref{plot_diffalpha}a). Now, AR and area can vary independently. We assume that the area constraint is satisfied and ignore the first term in Eq. (\ref{energyfunc}). This assumption is motivated by our simulation results, where the AR distribution is nearly independent of $\lambda_A$ (Fig. \ref{simcomp}f).

Next, consider the perimeter term: $\lambda_PP_i^2-2\lambda_PP_0P_i$; the first term represents contractility, and the second, effective adhesion; we have ignored the constant part. Exact description of adhesion is complex \cite{graner2017,hilgenfeldt2008,kafer2007}. The VM represents it as a line tension: $\sum_{\langle ij\rangle} \Lambda_{ij} \ell_{ij}$, where $\ell_{ij}$ is the length between two consecutive vertices, $i$ and $j$, and $\Lambda_{ij}$ gives the line tension. Since the degrees of freedom in the VM are the vertices, considering $\Lambda_{ij}$ constant, we obtain Eq. (\ref{energyfunc}). The constant $\Lambda_{ij}$ implies a regular cell perimeter between vertices. However, the entire cell boundary is in contact with other cells, and the perimeter is irregular in experiments \cite{kafer2007,atia2018}. We take a more general description, where the tension in a line-element $l_j$ (Fig. \ref{plot_diffalpha}a) is proportional to $l_j$ with strength $P_0$. Thus, the adhesion part becomes $-\lambda_P \tilde{K}P_0\sum_jl_j^2$, where $\tilde{K}$ is a constant.
On the other hand, the contractile term is proportional to $(\sum_j l_j)^2$. To make the calculation tractable, we use the Cauchy-Schwartz inequality \cite{arfkenweber} and write $(\sum_j l_j)^2\leq n\sum_j l_j^2=\nu \sum_jl_j^2$, where $\nu\sim \mathcal{O}(n)$ is a constant. Then, the perimeter part of $\mathcal{H}$ becomes $\nu \tilde{\lambda}_P\sum_j l_j^2=\nu\lambda_P(1-K P_0)\sum_jl_j^2$, where $K=\tilde{K}/\nu$. Thus, contractility and adhesion act as two competing effects (Fig. \ref{plot_diffalpha}a). Since it is unclear how to measure $P_0$ in experiments, we mainly restrict our discussions on $\lambda_P$ and $T$. Unless otherwise stated, we assume $P_0$ remains constant. However, we show in simulations that our mean-field theory captures the variation in $P_0$ (Fig. \ref{simcomp}h).
Cell division and apoptosis rates are usually low in epithelial monolayers \cite{poujade2007,park2015}; for example, they are of the order of $10^{-2}$ per hour and per day, respectively, for an MDCK monolayer \cite{puliafito2017}. Nevertheless, we show in the SM (Fig. S4), the general conclusions of the theory remain unchanged even when their rates are significant.

We describe the cell perimeter by the vector $\bx = \{x_1^1,x_1^2,x_2^1,x_2^2,\ldots x_n^1,x_n^2\}$, representing a set of $n$ points on the perimeter (Fig. \ref{plot_diffalpha}a). From the preceding discussion, we have $\mathcal{H}=\nu\tilde{\lambda}_P \bx(\mK\otimes\mathbf{I}_2)\bx'$, where $\bx'$ is the column vector, the transpose of $\bx$, $\mK$, the Kirchhoff's matrix \cite{eichinger1977,eichinger1980} with $\mK_{ii}=2$ and $\mK_{(i-1)i}=\mK_{i(i-1)}=-1$, and $\mathbf{I}_2$, the two-dimensional identity tensor. Thus, we have
\begin{equation}
\mathcal{H}/k_BT=\gamma \bx (\mK\otimes\mathbf{I}_2)\bx',
\end{equation}
where $\g=\nu\lambda_P(1-KP_0)/k_BT$, $k_B$ is the Boltzmann constant, and $T$, the temperature that parameterizes all possible activities, including the equilibrium temperature. To calculate $r$, we first obtain the moment of inertia tensor, a $2\times2$ tensor in spatial dimension two, in a coordinate system whose origin coincides with the center of mass (CoM) of the cell. Diagonalization of this tensor gives the two eigenvalues, $\Lambda_1$ and $\Lambda_2$, that are the squared-radii of gyrations around the respective principal axes. Therefore, $r=\sqrt{\Lambda_1/\Lambda_2}$, assuming $\Lambda_1\geq\Lambda_2$. However, a direct calculation of $\Lambda_1$ and $\Lambda_2$ is complex due to their anisotropic natures. It simplifies a bit for the radius of gyration, $s$, around the center of mass, and we have $s^2=\Lambda_1+\Lambda_2$ \cite{davis2018}. The distribution of $s^2$ is
\begin{equation}\label{distdef}
	P(s^2)=\frac{1}{Z}\int \prod_{\alpha=1}^2\delta(\sum_{j=1}^nx_j^\alpha)\delta(1-\frac{1}{ns^2}\bx\bx')e^{-\g \bx(\mK\otimes\mathbf{I}_2)\bx' }\frac{ \dot{x} }{ds}, 
\end{equation}
where $Z$ is the partition function, $\dot{x}=\prod_{\alpha=1}^2\prod_{j=1}^n dx_j^\alpha$, the volume element \cite{eichinger1977,eichinger1980}. Note that the definition of $P(s^2)$, Eq. (\ref{distdef}), assumes the Boltzmann distribution for perimeter at $T$ and accounts for all possible configurations allowed by $\mathcal{H}$. The applicability of an effective $T$ at the cellular level might be questionable. But the degrees of freedom here are much more microscopic (Fig. \ref{plot_diffalpha}a): within the CPM, it's a pixel, comparable to the resolution of the experimental image processing method. At this level of description, one can safely assume a Boltzmann distribution. Additionally, analyses of the experimental systems \cite{atia2018,kim2020} suggest the applicability of an effective $T$ description. The first $\delta$-function in Eq. (\ref{distdef}) ensures that the origin coincides with the CoM of the cell; the second $\delta$-function selects specific values of $s^2$, giving the distribution function. We next go to the normal-coordinate system that diagonalizes $\mK$. Integrating out the coordinates (see SM, Sec. S1), we obtain
\begin{equation}\label{distintform}
P(s^2)=\frac{|\mL_0|(\g n s^2)^{n-1}}{2\pi s} \int_{-\infty}^{\infty} \frac{e^{-i\beta} d\beta}{\prod_{j=1}^{n-1}\left(\g n s^2\lambda_j-i\beta\right)},
\end{equation}
where $\lambda_j$'s are the non-zero eigenvalues of $\mK$ whose diagonal form, without the zero-eigenvalue, is $\mL_0$. Taking contour integral of Eq. (\ref{distintform}), we obtain 
\begin{equation}
\label{finalexp}
P(s)=\frac{  (\g ns^2)^{n-1} }{2\pi s} |\mL_0| 2\pi i\sum_k Res(\lambda_k),
\end{equation}
where $\lambda_k$ are the distinct non-zero eigenvalues of $\mK$ and $Res(\lambda_k)$ gives the residue at the pole $\lambda_k$.
Since the cell perimeter must be closed-looped, $\mK$ is a tridiagonal matrix with periodicity. Therefore, the number of zero-eigenvalue must be one, and the lowest degeneracy of the non-zero eigenvalues must be two \cite{eichinger1977,kulkarni1999,*witt2009}; as explained below, the value of $k$ in $P(x,k)$ is determined by this second property. The leading-order contribution (SM, Sec. S1) comes from the smallest eigenvalue, $\lambda$, and is
\begin{equation}\label{dist_s2}
P(s^2) =C s^3\exp(-\tilde{\alpha} s^2),
\end{equation}
where $C$ is a normalization constant and $\tilde{\alpha}=\g n \lambda$. 
For cells, the two eigenvalues, $\L_1$ and $\L_2$, are not independent since $\sqrt{\L_1\L_2}=A$, the cell area. Using this relation, we obtain $s^2=A(r+1/r)$ and from Eq. (\ref{dist_s2}), we obtain
\begin{equation}\label{distAR}
P(r)=\frac{1}{\mathcal{N}}\left(r+\frac{1}{r}\right)^{3/2}\left(1-\frac{1}{r^2}\right)e^{-\alpha(r+\frac{1}{r})},
\end{equation}
where $\mathcal{N}$ is the normalization constant: $\mathcal{N} = \Gamma(5/2)/\alpha^{5/2}-W(5/2)  _1F_1(5/2,7/2,-2\alpha)$, where $W(x)=2^x/x$, and $_1F_1(a,b,c)$ is the Kummer's confluent Hypergeometric function \cite{erdelyibook},
$\alpha \propto \lambda_P(1-KP_0)/T$. 
As detailed in the SM (Sec. S2), Eq. (\ref{dist_s2}), together with the constraint of confluency \cite{weaire1986,gezer2021}, give the distribution for the scaled area $a=A/\bar{A}$, where $\bar{A}$ is the average of area, $A$. It is a Gamma distribution, with a single parameter $\mu$,
\begin{equation}\label{distarea}
P(a)=\frac{\mu^\mu}{\Gamma(\mu)} a^{\mu-1}\exp[-\mu a].
\end{equation}
Since $\mu$ is related to the constraint of confluency, it should be independent of $\lambda_P$; our simulations show that this is indeed true (Fig. \ref{simcomp}f). Therefore, $\alpha$ and $\mu$ together can distinguish how the model parameters $\lambda_P$ and $T$ are affected by changing conditions such as maturation.
Equation (\ref{distAR}) provides a remarkable description, where all the microscopic details of the system enter through a single parameter, $\alpha$; it has profound implications leading to the universal behavior as we now illustrate.


\subsection*{Aspects of universality}

We first show the PDF of the aspect ratio, $P(r)$, at different values of $\alpha$ in Fig. \ref{plot_diffalpha}(b), $P(r)$ decays faster as $\alpha$ increases, as expected from Eq. (\ref{distAR}). The plots look remarkably similar to the experimental results shown in Ref. \cite{atia2018}; we present detailed comparisons with experiments later. 
Reference \cite{atia2018} has demonstrated that the PDFs of the scaled aspect ratio, $r_s=(r-1)/(\bar{r}-1)$, where $\rav$ is the ensemble-averaged $r$, across different systems follow a near-universal behavior. We now plot the PDFs of $r_s$ in Fig. \ref{plot_diffalpha}(c). The PDFs {\em almost} overlap, but they are not identical. 
A closer look at Eq. (\ref{distAR}) shows that if $\rav+1/\rav$ goes as $1/\alpha$, we can scale $\alpha$ out of the equation and obtain a universal scaled distribution. However, as shown in Fig. \ref{plot_diffalpha}(d), there is a slight deviation in $\rav+1/\rav$ with the functional form of $1/\alpha$. As shown in the inset of Fig. \ref{plot_diffalpha}(d), $\rav$ also slightly deviates from the $1/\alpha$ form.
This tiny deviation implies that the PDF of $r_s$ is not universal. If one ignores $1/r$ compared to $r$, Eq. (\ref{distAR}) becomes a $k$-Gamma function for $r_s$ with $k=2.5$. However, since $r\sim\mathcal{O}(1)$, this cannot be a good approximation, and the observed spread of $k$ around 2.5 is natural when fitted with this function \cite{atia2018,li2018,kim2020,wenzel2021}.
On the other hand, since the deviation (Fig. \ref{plot_diffalpha}d) is minute, the PDFs of $r_s$ for different systems look {\em virtually} universal. This result is a strong prediction of the theory and, as we show below, is corroborated by available experimental data on diverse epithelial systems.

Although the PDFs of $r_s$ are not strictly universal, there is another aspect, $sd$ vs $\rav$, which is universal.
We show the $sd=\sqrt{\overline{r^2}-\rav^2}$ as a function of $\rav$ in Fig. \ref{plot_diffalpha}(e). Since there is only one parameter, $\alpha$, in Eq. (\ref{distAR}), it determines both $sd$ and $\rav$. The monotonic dependence of $\rav$ on $\alpha$ (inset, Fig. \ref{plot_diffalpha}(d)) implies a unique relationship between them. Therefore, we can express $\alpha$ in terms of $\rav$ and, in turn, $sd$ as a function of $\rav$. Since there is no other system-dependent parameter in this relation, it must be universal. Note that $\alpha\propto \lambda_P/T$ at a constant $P_0$, thus, $\alpha$ increases as $\lambda_P$ increases or $T$ decreases. Both $sd$ and $\rav$ become smaller as $\alpha$ increases, and the system on the $sd$ vs $\rav$ plot moves towards the origin (Fig. \ref{plot_diffalpha}e).
From the perspective of the dynamics, the relaxation time, $\tau$, of the system grows as $\alpha$ increases \cite{souvik2021}. Thus, small $\rav$ and large $\tau$, that is, less elongated cells and slow dynamics, follow each other, and the energy function, Eq. (\ref{energyfunc}), controls both behaviors. Finally, we show some representative PDFs, Eq. (\ref{distarea}), at different values of $k$ for the scaled area $a=A/\bar{A}$. The PDF of $a$ has been argued to be universal \cite{wilk2014}; our theory shows that although they follow the same distribution, they need not be identical.

\subsection*{Comparison with simulations}
We now compare our analytical theory with simulations of two distinct confluent models: the CPM and the VM. Figures \ref{simcomp}(a) and (b) show representative plots for the comparison of the PDFs of $r$ within the CPM, and the VM, respectively, where the lines represent fits with Eq. (\ref{distAR}). Figure \ref{simcomp}(a) shows data with varying $\lambda_P$, and Fig. \ref{simcomp}(b) shows data with changing $T$. As discussed above, our theory predicts almost universal behavior for the PDFs of $r_s$ (Fig. \ref{plot_diffalpha}c). We plot the simulation data for the PDFs of $r_s$ for both the models at different parameters in Fig. \ref{simcomp}(c); the PDFs almost overlap with each other, consistent with the theory (see SM, Fig. S2 for more results). Within our mean-field theory, we have ignored the area term in Eq. (\ref{energyfunc}), arguing that the perimeter term contributes dominantly to the shape. We have obtained $\alpha$ at fixed $\lambda_P$, $P_0$, and $T$ but varying $\lambda_A$. As shown in the insets of Figs. \ref{simcomp}(a) and (b), within both the CPM and the VM, $\alpha$ remains almost constant with varying $\lambda_A$, justifying our assumption. An important prediction of the theory is that the parameter $\alpha$, which governs the behavior of the cell shape variability, is linearly proportional to both $\lambda_P$ and $1/T$, hence with $\lambda_P/T$. This prediction also agrees with our simulations (Fig. \ref{simcomp}d and SM, Fig. S6). The slopes within the CPM and the VM are different; this possibly comes from the distinctive natures of the two models, but the qualitative behaviors are the same.

We now show the behavior of the PDF of scaled cell area, $a$, in Fig. \ref{simcomp}(e) with varying $\lambda_A$. The lines represent the fits with Eq. (\ref{distarea}). Larger values of $\lambda_A$ ensure the area constraint is more effective and the distributions become sharply peaked around the average area. The inset of Fig. \ref{simcomp}(e) shows that $\mu$ almost linearly increases with $\lambda_A$. 
Since the area term is related to the cell height that remains nearly constant and the geometric constraint of confluency, we don't expect a substantial variation in $\lambda_A$ in a particular system. However, $\mu$ also varies with $T$ (see SM, Fig. S7). Thus, in contrast to what has been proposed \cite{wilk2014}, as $T$ changes, the PDF of $a$, though well-described by a single parameter $\mu$ via Eq. (\ref{distarea}), can still be different.

Figure \ref{simcomp}(f) shows $\alpha$ vs $\mu$ within the CPM when we vary one of the parameters, $\lambda_A$, $\lambda_P$, and $T$, keeping the other two fixed. First, when $\lambda_A$ increases, the value of $\mu$ increases, but $\alpha$ remains almost constant (also see insets of Fig. \ref{simcomp}a and b). Next, when $\lambda_P$ increases, although $\alpha$ linearly increases, $\mu$ remains nearly the same. Finally, both parameters linearly increase with $1/T$; since higher $T$ implies more fluctuations, decreasing $T$ helps both $r$ and $a$ to become sharply peaked (see SM, Figs. S6, S7 for their specific behaviors, and results within the VM). These results show when $\lambda_A$ remains constant, varying $\lambda_P$ and $T$ have distinctive effects on $\mu$ and $\alpha$. 
These results are significant from at least two aspects: First, $\mu$ comes from the constraint of confluency (see SM, Sec. S2), which should depend only on the area and be independent of the perimeter. Thus, the $\lambda_P$-independence of $\mu$ validates the phenomenological implementation \cite{weaire1986,gezer2021} of this constraint.
Second, these results can provide crucial insights regarding the model parameters. Maturation of a monolayer can affect both $\lambda_P$ and $T$. Additional junctional proteins may be employed during maturation to increase $\lambda_P$. On the other hand, different forms of activity may reduce decreasing $T$. Since $\alpha$ increases linearly with $\lambda_P/T$, AR alone is not enough to determine the dominant mechanism during the maturation process. However, assuming that $\lambda_A$ remains constant in a particular system, simultaneous measurements of $\mu$ and $\alpha$ allow distinguishing effects of changing physical conditions, such as maturation, on the individual parameters.

\begin{figure*}
	\centering
	\includegraphics[width=0.8\textwidth]{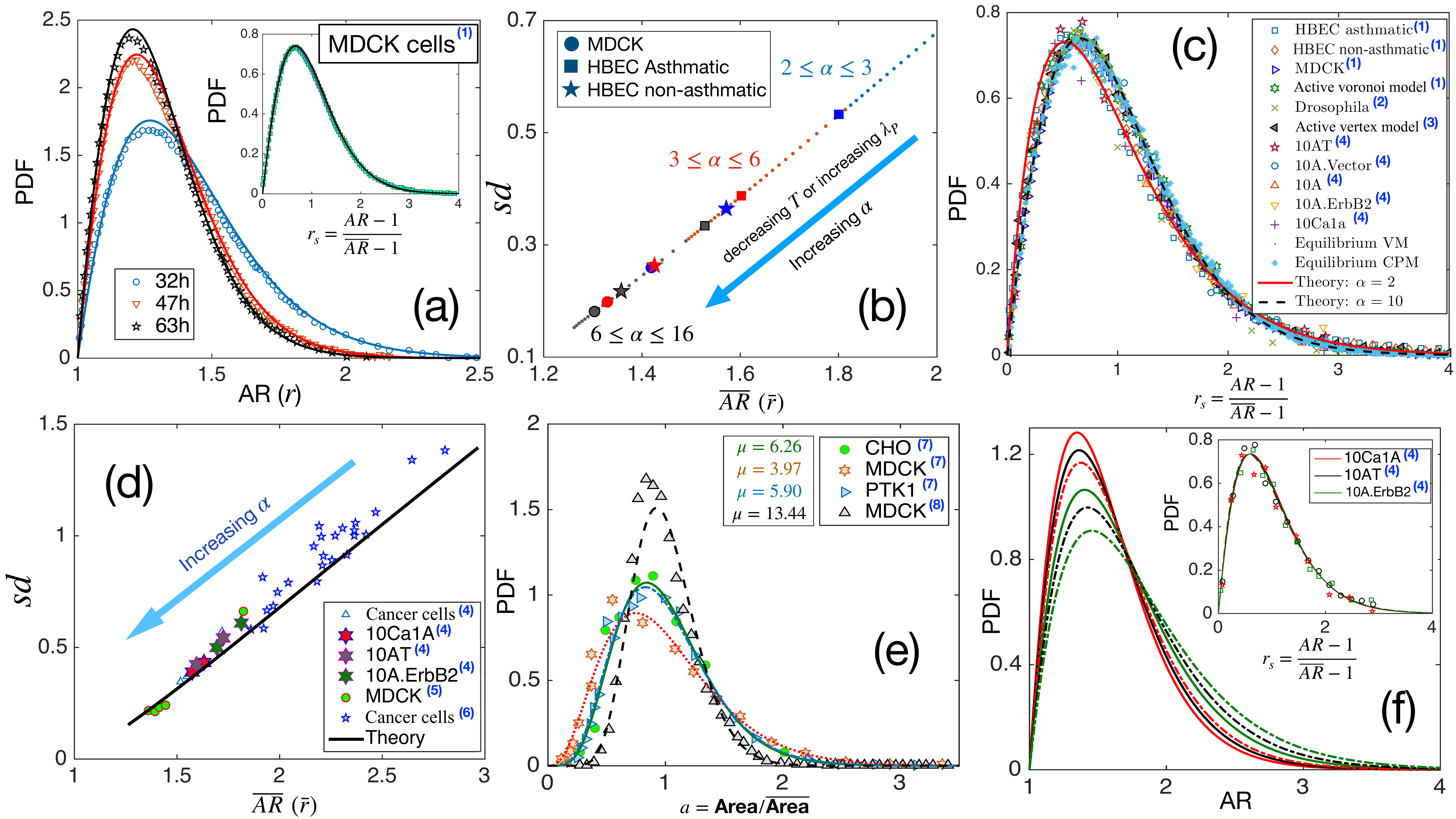}
	\caption{Comparison with existing experiments. (a) PDF for AR. Symbols are experimental data at three different times for MDCK cells, and lines represent the fits with our theory, Eq. (\ref{distAR}). The values of $\alpha$ are quoted in Table \ref{alphavalues}. {\bf Inset:} PDF of the $r_s$, the lines are theory, and the symbols are data. (b) Using the values of $\alpha$, obtained for the three sets of data as quoted in Table \ref{alphavalues}, we obtain $sd$ vs $\rav$ using our theory. Different symbols represent the types of the systems, and the colors blue, red, and black represent early, intermediate, and later time data, respectively. With maturation, the system moves towards lower $\rav$ and smaller $sd$. (c) PDF for $r_s$ for a wide variety of systems, shown in the figure, seems to be virtually universal, consistent with our theory. (d) The theory predicts a universal relation for $sd$ vs $\rav$. Symbols are data for different systems, and the line is our theoretical prediction. (e) PDF of the scaled area for different epithelial systems and the lines represent the fits with Eq. (\ref{distarea}). (f) Predictive power of the theory: we use the $\rav$ for the three sets of cells from Ref. \cite{kim2020}, as marked by the hexagons in (d), and obtain the corresponding values of $\alpha$ and obtain the PDFs for $r$ using Eq. (\ref{distAR}). The colors correspond to the type of cells in (d), and the continuous line corresponds to the lower $\rav$ data. {\bf Inset:} Lines are the theoretical PDFs for $r_s$ using the values of $\alpha$ for different cells, and the symbols show the experimental data. We have collected the experimental and some of the simulation data from different papers. Data taken from other papers are marked with a blue superscript in the legends. The sources are as follows: (1) \cite{atia2018}, (2) \cite{li2020}, (3) \cite{lin2018}, (4) \cite{kim2020}, (5) \cite{fujii2019}, (6) \cite{ilina2020}, (7) \cite{wilk2014} and (8) \cite{puliafito2017}.}
	\label{exptcomp}
\end{figure*}

We next verify the universal result of the theory: $sd$ vs $\rav$. Figure \ref{simcomp}(g) shows $sd$ vs $\rav$ within the CPM; we plot the theoretical prediction by the dotted line for comparison. The theory predicts that the state points move towards the origin as $\alpha$ increases (Fig. \ref{plot_diffalpha}e); this is consistent with our simulations.
Since $\alpha\propto \lambda_P/T$, higher $\alpha$ should correspond to slower dynamics. To test this hypothesis, we have simulated the CPM at different $\lambda_P$, $P_0$, and $T$ to obtain the relaxation time, $\tau$ (see SM, Sec. S3 for details). From these control parameters, we have calculated $\alpha$ and then $\rav$ using our theory. We show $\tau$ as functions of $\alpha$ and $\rav$ in the inset of Fig. \ref{simcomp}(g); it is clear that indeed $\tau$ grows as $\alpha$ increases or $\rav$ decreases. To show this behavior of $\tau$ on the same plot as $sd$ vs $\rav$, we plot $2.5/\ln\tau-0.25$ in the main-figure as a function of $\rav$.
A monolayer fluidizes under compressive or stretching experiments, where cell shape changes, but not cell area \cite{krishnan2012,park2015,atia2018}. Such perturbations make the cells more elongated, increasing $\rav$; thus, our theory rationalizes the decrease in $\tau$ associated with fluidization. Finally, we show that our mean-field result that $\alpha$ decreases linearly with $P_0$ agrees with simulations (Fig. \ref{simcomp}h).
Further, to test our hypothesis that our main results remain unchanged in the presence of cell division and apoptosis,
we have simulated the CPM, including these processes. As shown in the SM, Sec. S7, the simulation results justify our hypothesis.

\subsection*{Comparison with existing experiments}

Having shown that our theory agrees well with the simulations, both within the CPM and the VM, we next confront it with the existing experimental data. We first compare the theory with data taken from Ref. \cite{atia2018} for three different confluent cell monolayers: the MDCK cells, the asthmatic HBEC, and the nonasthmatic HBEC. We chose the PDFs at three different times from Figs. 3 (a) and (c) in Ref. \cite{atia2018}. We fit Eq. (\ref{distAR}) with the data to obtain $\alpha$ and present their values in Table \ref{alphavalues}; the corresponding fits for the MDCK cells are shown in Fig. \ref{exptcomp} (a)  (see SM, Fig. S8 for the other fits). Table \ref{alphavalues} shows that $\alpha$ increases with maturation. Thus, progressive maturation can be interpreted as an increase in either $\lambda_P$ or $1/T$ or both. The PDFs for $r_s$ corresponding to the MDCK cells are shown in the inset of Fig. \ref{exptcomp}(a) together with one set of experimental data \cite{atia2018}; note that this is not a fit, yet the theory agrees remarkably well with the data.

We next calculate $sd$ as a function of $\rav$ using the values of $\alpha$, noted in Table \ref{alphavalues} for the three systems. They are shown in Fig. \ref{exptcomp}(b) along with the theory prediction. With maturation, the state points move towards lower $\rav$, represented by the arrow in Fig. \ref{exptcomp}(b). As shown in Fig. \ref{simcomp}(g), larger $\alpha$ corresponds to a system with higher $\tau$. Thus, with maturation, as the PDFs become sharply peaked, as the cells look more roundish and $\rav$ becomes smaller, the system becomes more sluggish. This effect of maturation is the same in all the systems (Fig. \ref{exptcomp}b) and agrees with the interpretation presented in Ref. \cite{atia2018}. We have also examined that the theoretical prediction of $sd$ vs $\rav$ agrees well with the experimentally measured values shown in Ref. \cite{atia2018}.

\begin{table}
	\caption{Values of $\alpha$ from fits of Eq. (\ref{distAR}) with the PDFs of $r$. The data for the three systems are taken as function of maturation time, in units of h (hours) and d (days). The experimental data are taken from Ref. \cite{atia2018}.}
	\label{alphavalues}
	\vspace{0.1cm}
	\begin{tabular}{C{3.5cm}| C {1.5cm}| C{2cm}}
		\hline
		Cell type & Time & Value of $\alpha$\\
		\hline
		\, & 32 h & 7.535 \\
		MDCK & 47 h & 10.998 \\
		\,  & 63 h & 12.496 \\
		\hline
		\,   & 6 d & 3.060 \\
		Asthmatic HBEC & 14 d & 4.472 \\
		\,  & 20 d & 5.389\\
		\hline
		\, & 6 d & 4.798\\
		Non-asthmatic HBEC  & 14 d & 7.358 \\
		\,  & 20 d & 9.602\\
		\hline
	\end{tabular}
\end{table}

Our theory predicts that the PDF for $r_s$, though not strictly universal, should be almost the same for different systems (Fig. \ref{plot_diffalpha}c). This prediction is a consequence of a crucial aspect of the theory: all the system-specific details enter via a single parameter, $\alpha$ in Eq. (\ref{distAR}). As shown in Fig. \ref{plot_diffalpha}(d), $\rav+1/\rav$ deviates slightly from the behavior $1/\alpha$. This slight deviation implies that the PDF for $r_s$ can not be strictly universal and manifests as a variation in $k$ when the PDF is fitted with the $k$-Gamma function in different experiments and simulations \cite{atia2018,kim2020,li2018,lin2020,wenzel2021}. Nevertheless, since the deviation in Fig. \ref{plot_diffalpha}(d) is very weak, the values of $k$ are very close to each other. Therefore, the PDFs for $r_s$ for diverse epithelial systems--in experiments, simulations, and theory--should be virtually universal. To test this prediction, we have collected existing experimental and simulation data on different systems and show the PDFs of $r_s$ on the same plot in Fig. \ref{exptcomp}(c). The variety in our chosen set is spectacular: it consists of different cancer cell lines \cite{kim2020}, both asthmatic and non-asthmatic HBEC cells, MDCK cells \cite{atia2018}, Drosophila wing disk \cite{lin2020}, simulations data on both active \cite{li2018} and equilibrium versions of the VM, the active Voronoi model \cite{atia2018}, and the CPM. Yet, the PDFs, as shown in Fig. \ref{exptcomp}(c), look virtually universal and in agreement with our analytical theory.

Additionally, our theory predicts a strictly universal $sd$ vs $\rav$. Since this relation does not have any system-specific details, data across diverse confluent monolayers must follow a universal relationship. We have collected existing experimental data for several systems: cancerous cell lines \cite{kim2020}, human breast cancer cells \cite{ilina2020}, and a jammed epithelial monolayer of MDCK cells \cite{fujii2019}. Figure \ref{exptcomp}(d) shows the experimental data together with our theoretical prediction; the agreement with our theory, along with the aspect of universality, is truly remarkable. As $\alpha$ increases, dynamics slows down, and the points on this plot move towards lower $\rav$. This result is consistent with the finding that cell shapes are more elongated and variable as the dynamics become faster in different epithelial systems \cite{atia2018,park2015}.

We have argued that simultaneous measurements of the PDFs of cell area and AR distinguish the effects of maturation on the two key parameters: $\lambda_P$ and $T$. The argument relies on the negligible effect of the perimeter constraint on $\mu$ (Eq. \ref{distarea}). We now show a comparison of our theoretical result for the PDF of $a$ with existing experiments. Figure \ref{exptcomp}(e) shows experimental data for four different systems \cite{wilk2014,puliafito2017} and the corresponding fits of Eq. (\ref{distarea}). Unlike what has been proposed, that epithelial monolayers have a universal area distribution \cite{wilk2014}, we find, in agreement with experiments, the distribution can vary, although the functional form remains the same.

What are the implications of these universal aspects of cell shape variability and our theory? Cell shape controls several crucial biological functions such as the mitotic-orientation \cite{wyatt2015,bosveld2016,xiong2014} and cell fate \cite{mcbeath2004,wang2011,roskelley1994}. Our theory shows that the microscopic system properties are encoded via a single parameter, $\alpha$. Consequently, knowledge of one of the observables, such as $\rav$, contains the information of the entire statistical properties in a monolayer. We now illustrate this predictive aspect of the theory. Experimental measurement of an average property is usually less complex and more reliable. We have collected the data for $\rav$ from the supplementary material of Ref. \cite{kim2020} for three different systems: 10Ca1A, 10AT, and 10A.ErbB2, shown by the hexagons in Fig. \ref{exptcomp}(d). From these average values, we obtain $\alpha$, which we use to theoretically calculate the PDFs for $r$, as shown in Fig. \ref{exptcomp}(f). The inset of Fig. \ref{exptcomp}(f) shows our theoretical PDFs for $r_s$, together with the corresponding experimental data for comparison. The excellent agreement demonstrates that cell shape variability results from the geometric constraint imposed by the energy function, Eq. (\ref{energyfunc}), and is not a choice but inevitable for such systems. This result, we believe, will foster analysis of diverse epithelial systems to understand the interrelation between geometric properties and biological functions within a unified framework.

\subsection*{Discussion}
We have obtained a mean-field theory for cell shape variability through the energy function $\mathcal{H}$, Eq. (\ref{energyfunc}) \cite{farhadifar2007,honda1978,bi2015,bi2016,souvik2021}. The geometric restriction of confluency is a strong constraint on cell area. Considering that it is satisfied and that the cell cortex, described by the perimeter term, is crucial in determining the cell shape, allowed us to ignore the area term and obtain the distribution for AR. Recent experiments and simulations have shown that cell shape variability is virtually universal in a confluent epithelial monolayer \cite{atia2018,wenzel2021,kim2020,ilina2020}; our work provides the theoretical basis for such behavior. 
Contrary to common beliefs \cite{atia2018,kim2020}, the jamming transition is not related to the cell shape distribution. We have analytically obtained the functional form for cell shape variability and find that the microscopic system properties enter the distribution via a single parameter, $\alpha$: this leads to the universal behavior for $sd$ vs $\rav$ and a virtually universal distribution for the scaled aspect ratio $r_s$. Our theoretical results are verified in simulations within the CPM, on the square and hexagonal lattices, and the VM. They also agree remarkably well with the existing experiments. Thus far, the PDF of $r_s$ has been fitted with a $k$-Gamma function with $k$ being around 2.5. We show that the slight variation in $k$ comes from the fact that the PDF is not strictly but almost universal. On the other hand, $k\simeq 2.5$ is a direct consequence of a mathematical property: the lowest degeneracy of the eigenvalues of the connectivity matrix being two for a closed-looped object, here the perimeter.

A better understanding of the connection between the theoretical parameters and different system properties is crucial to exploit the universal aspects for deeper insights. Since all of the parameters combine into $\alpha$, the effects of changing physical conditions on the individual parameters are difficult to determine from the measurements of AR alone. Within our theory, $\lambda_P$ describes the cortical properties, and $T$ parameterizes different biological activities, including temperature. These are effective parameters, and their direct measurement in biological systems is impractical. Our theory provides an indirect way to estimate these parameters. The cell area in a confluent system is geometrically constrained. We have used the phenomenological implementation of the constraint of confluency and obtained the PDF for $a$ \cite{weaire1986}. It is a Gamma function, described by a single parameter $\mu$. Reference \cite{wilk2014} has proposed that the PDF of $a$ is universal in various epithelial monolayers. However, we show that though the area distribution follows the same function, there is a variation in $\mu$. Our work connects $\mu$ to the microscopic model parameters of Eq. \ref{energyfunc}. In particular, $\mu$ should be independent of $\lambda_P$, whereas $\alpha$ varies linearly with both $\lambda_P$ and $1/T$. This distinction, assuming $\lambda_A$ remains constant, allows inferring the effects of maturation on the individual model parameters. 
We have neglected cell division, growth, and apoptosis in our theory. However, the consequences of these processes, and the effects of chemical gradients \cite{kinjal2018}, should also enter the distribution via $\alpha$, as the algebraic part comes from the topological property that remains the same. We have included them in our simulations, and the results support this expectation (SM, Sec. S7); this implies that the main predictions of the theory remain valid even in the presence of these processes.

Our work demonstrates that a single parameter describes both the cell shape statistics and the dynamics.
We have shown in our simulations that the relaxation time, $\tau$, grows as $\alpha$ increases or $\rav$ decreases. Experiments on confluent cellular monolayer have also reported similar results \cite{atia2018,park2015,kim2020}. Thus, as cells become more compact and their shape variability reduces, the monolayer becomes more sluggish. This result may have far-reaching consequences. Most experiments usually measure $\rav$ and analyze biological functions via $\rav$ \cite{wyatt2015,bosveld2016}. Our theory implies that such knowledge contains a wealth of information. One can obtain $\alpha$ from $\rav$, and all other properties, such as the distribution, the standard deviation, and the dynamics, can be analytically calculated. Whether this correlation between the statics and dynamics persists in all regimes, including the glassy-regime, remains an open question. 
Additionally, in biology, it is well-recognized that different levels of organizations are mechanistically related. One fundamental open question is how molecular-level events are related to cellular machines that control the cell shape \cite{gilmour2017}. The striking predictability, demonstrated by our theory, where $\rav$ determines the PDF, shows that the statistical distribution of cell shape is unavoidable. How do different cells respond to this inevitable distribution? Is cellular response similar across diverse systems? How is it related to organ-level morphogenesis? Having a single parameter that describes the static and dynamic aspects at the cellular level should help to compare and analyze different systems and answer these questions.

In conclusion, we have developed a mean-field theory, considering cortical contractility and adhesion are crucial in determining the shape. We have analytically derived the cell shape variability, characterized via the AR. The PDF for AR is described by a single parameter, $\alpha$. As a result, $sd$ vs $\rav$ becomes universal, and the PDF for the scaled aspect ratio, $r_s$, is virtually universal. The analytical form for the PDF of $r_s$ can be roughly approximated as a $k$-Gamma distribution \cite{aste2008} that has been fitted with the existing experimental data \cite{atia2018,kim2020}; however, the distribution is unrelated to the jamming transition, in contrast to the general belief. The near-universal value of $k \sim 2.5$ is the consequence of a mathematical property, and the variation results from the fact that the PDFs are not strictly universal. $\alpha$ can also provide information on dynamics. In addition, simultaneous measurements of the PDFs of $a$ and $r$ can relate the model parameters with the changing physical properties. Such a relationship should further strengthen the link between the energy function, Eq. (\ref{energyfunc}), and an epithelial monolayer. Having a single parameter for the statistical and dynamical aspects of an epithelial monolayer should foster a detailed comparison of diverse systems, help to analyze the relation of biological functions to shapes, and illuminate the mechanisms of cellular response to the inevitable shape variability.

\subsection*{Acknowledgements}
We thank Mustansir Barma and Sumedha for many enlightening discussions and Sam. A Safran for comments. We acknowledge support of the Department of Atomic Energy, Government of India, under Project Identification No. RTI 4007.

\bibliography{aspectratioref.bib}

\end{document}


\title{Supplementary Material : The origin of universal cell shape variability in a confluent epithelial monolayer}
\author{Souvik Sadhukhan}
\affiliation{TIFR Centre for Interdisciplinary Sciences, Tata Institute of Fundamental Research, Hyderabad - 500046, India}
\author{Saroj Kumar Nandi}
\affiliation{TIFR Centre for Interdisciplinary Sciences, Tata Institute of Fundamental Research, Hyderabad - 500046, India}

\maketitle

In this supplementary material, we first provide the details of the analytical derivation of the probability distribution functions (PDFs) of the aspect ratio (AR) in Sec. \ref{ARderivation}, and the scaled area, $a$, in Sec. \ref{areaderivation}. We next provide the details of the simulations in Sec. \ref{simdetails}, and explain the procedure to calculate the AR in our simulations in Sec. \ref{ARprocedure}. We then present data of the CPM on the hexagonal lattice in Sec. \ref{hexresults}, and show additional data in Sec. \ref{lamAAR}, supporting the claim that AR does not depend on $\lambda_A$. Section \ref{divapop} shows simulation results when we include cell division and apoptosis. We present the PDF of $a$ within the vertex model in Sec. \ref{areadistVM}, show the dependence of the parameter $\alpha$ on $\lambda_P$ and $T$ in Sec. \ref{alpha_individual}. The $T$-dependence of $\mu$ is shown in Sec. \ref{muonT} and, finally, the fits with the experimental data of HBEC cells are shown in Sec. \ref{hbecdatafit}.

\renewcommand{\theequation}{S\arabic{equation}}
\renewcommand{\thesubsection}{S\arabic{subsection}}
\renewcommand{\thefigure}{S\arabic{figure}}

\subsection{Details of the derivation}
\label{ARderivation}
As shown in the main text, defining $\bx=\{ x_1^1,x_1^2,x_2^1,x_2^2, \ldots x_n^1,x_n^2\}$ as a particular configuration of the perimeter of a specific cell, we can write the energy in units of $k_BT$ for this cell as
\begin{equation}
\mathcal{H}/k_BT=\g \bx (\mK\otimes \mathbf{I}_2)\bx'
\end{equation}
where $k_BT$ is Boltzmann constant times temperature, $\g=\nu\lambda_P(1-KP_0)/k_BT$ , $\mK$ is the $n$-dimensional Kirchoff's matrix with $\mK_{ii}=2$ and $\mK_{(i-1)i}=\mK_{i(i-1)}=-1$. $\mathbf{I}_2$ is the two-dimensional identity tensor and $\otimes$ denotes the tensor product. $\bx'$ is a column vector, the transpose of $\bx$. Then, the distribution of the radius of gyration \cite{eichinger1977,eichinger1980} can be written as
\begin{equation}\label{distdef}
P(s^2)=\frac{1}{Z}\int \prod_{\alpha=1}^2\delta(\sum_{j=1}^nx_j^\alpha)\delta(1-\frac{1}{ns^2}\bx\bx')\exp(-\g \bx(\mK\otimes\mathbf{I}_2)\bx' )\frac{ \dot{x} }{ds}
\end{equation}
where the volume element $\dot{x}$ is defined as $\dot{x}=\prod_{\alpha=1}^2\prod_{j=1}^n dx_j^\alpha$. The first $\delta$-function in Eq. (\ref{distdef}) coincides center of mass with the origin of the coordinate system. The squared radius of gyration is $s^2=n^{-1}\bx\bx'$ and the second $\delta$-function in Eq. (\ref{distdef}) ensures that the correct values of $s$ are chosen for the distribution. $Z$ is the partition function of the system. Since the radius of gyration does not depend on the coordinate system, we are allowed to chose one that diagonalizes $\mK$. Say the diagonal matrix is $\mL$, and $\mq$ represents the normal coordinates in this system.

The radius of gyration can be defined as the root-mean-square distance of different parts of a system either from its center of mass or around a given axis. We have designated the former as $s$, defined as
\begin{equation}
s=\sqrt{\frac{1}{N}\sum_{i=1}^N (\bx_i-\bx_{CM})^2},
\end{equation}
where $N$ is the total volume element in the system with coordinates $\bx_i$, and $\bx_{CM}$ is the center of mass (CoM) of the system. The other two radii of gyration can be defined around the two principal axes (since we are in spatial dimension two) passing through the CoM. We calculate these two radii of gyration by writing the inertia tensor in a coordinate system whose origin coincides with the CoM and diagonalizing the tensor. The eigenvalues, $\L_1$ and $\L_2$, give the square of the radii of gyration. Thus, the aspect ratio, $r$, is obtained as $r=\sqrt{\L_1}/\sqrt{\L_2}$. Due to the anisotropic nature of $\L_1$ and $\L_2$, a direct calculation for their distributions is more complex than that of $s$. We first calculate the distribution for $s$ and then, using this result, obtain the distribution of $r$.

Equation (\ref{distdef}) can be written in the normal coordinate system as
\begin{equation}
P(s^2)=\frac{1}{Z}\int \prod_{\alpha=1}^2\delta( q_n^\alpha)\delta(1-\frac{1}{ns^2}\mq\mq')\exp(-\g \mq(\mL\otimes\mathbf{I}_2)\mq')\frac{\dot{q}}{ds}
\end{equation}
where we have used $q_n^\alpha\propto \sum x_j^\alpha$ that corresponds to the zero-eigenvalue mode of the matrix. Integrating over $q_n^\alpha$, we get rid of this zero-eigenvalue that gives translation. Thus,
\begin{equation}\label{Pofsdef}
P(s^2)=\frac{1}{Z}\int \delta(1-\frac{1}{ns^2}\mq_0\mq'_0) \exp(-\g \mq_0(\mL_0\otimes\mathbf{I})\mq_0')\frac{\dot{q}_0}{ds}
\end{equation}
where we have defined $\mq_0$ as the $2(n-1)$ dimensional vector excluding the coordinates corresponding to the  zero-eigenvalue. The normalization factor, $Z$, can be calculated exactly through the integration as
\begin{equation}
Z\equiv\int  \exp(-\g \mq_0(\mL_0\otimes\mathbf{I}_2)\mq_0')\dot{q}_0=\left(\frac{\pi}{\g}\right)^{(n-1)}|\mL_0|^{-1}.
\end{equation}
Note that the integration in the calculation of $P(s^2)$ is around the boundary of the cell; to separate out the radial part, we now write the volume element in polar coordinate $\mbu$: $\mq_0=n^{1/2}s\mbu$. Then $\dot{q}_0=n^{(n-1)}s^{2(n-1)-1}ds\dot{u}$ and $\mq_0\mq_0'/ns^2=\mbu\mbu'=1$. Thus, we obtain from Eq. (\ref{Pofsdef})
\begin{equation}
P(s^2)=\mathcal{A}\int_{-\infty}^{\infty}d\beta\int e^{-i\beta}e^{-\g ns^2 \mbu[(\mL_0-\frac{i\beta}{n\g s^2}\mathbf{I}_{n-1})\otimes\mathbf{I}_2]\mbu'}\dot{u} ,
\end{equation}
with $\mathcal{A} = \left(\frac{\g}{\pi}\right)^{(n-1)}|\mL_0|\frac{1}{2\pi}n^{(n-1)}s^{2(n-1)-1}$. $\mathbf{I}_{n-1}$ is the identity matrix of rank $n-1$. Carrying out the integration over $\mbu$, we obtain 
\begin{equation}
P(s^2)=\mathcal{A}\int_{-\infty}^\infty d\beta e^{-i\beta}\frac{\left(\frac{\pi}{\g ns^2}\right)^{(n-1)}}{\big| \mL_0-\frac{i\beta}{\g ns^2}\mathbf{I} \big|}.
\end{equation}
Using the value of $\mathcal{A}$, we obtain
\begin{align}\label{distintform}
P(s^2)&=\frac{1}{2\pi s}\int_{-\infty}^{\infty}d\beta \frac{e^{-i\beta} }{\prod_{j=1}^{n-1}\left(1-\frac{i\beta}{\g n s^2\lambda_j}\right)} \nonumber\\
 &=\frac{(\g n s^2)^{n-1}}{2\pi s}|\mL_0|\int_{-\infty}^{\infty}d\beta  \frac{e^{-i\beta}}{\prod_{j=1}^{n-1}\left(\g ns^2\lambda_j-i\beta\right)},
\end{align}
where $\lambda_j$'s are the eigenvalues of $\mK$. The integral in Eq. (\ref{distintform}) can be performed via the contour integral and the resultant solution can be written as
\begin{equation}
\label{finalexp}
P(s^2)=\frac{  (\g ns^2)^{n-1} }{2\pi s} |\mL_0| 2\pi i\sum_k Res(\lambda_k),
\end{equation}
where $\lambda_k$ are the distinct eigenvalues of $\mK$ and $Res(\lambda_k)$ gives the residue at the pole $\lambda_k$. As we show below, the residues will have a term $\exp[-n\g s^2\lambda_k]$ and in the limit $s^2\to\infty$, only the smallest $\lambda_k$ will contribute.

Since the cell perimeter must be closed-looped, $\mK$ is a tridiagonal matrix with periodicity. Therefore, the number of zero-eigenvalue must be one, and the lowest degeneracy of the non-zero eigenvalues must be two \cite{kulkarni1999,witt2009,eichinger1977,eichinger1980}. We have already integrated out the coordinate corresponding to the zero-eigenvalue. Let us designate the lowest non-zero eigenvalue as $\lambda$. The pole corresponding to $\lambda$ is located at $\beta = -i\g n s^2 \lambda$, and of order 2. Thus, we obtain the residue as
\begin{align}
\label{Residue}
\text{Res}= \frac{d}{d\beta}\Big [ \frac{ e^{-i\beta }}{\prod_{j=1}^{n-3}\left({\g n s^2\lambda_j}-{i\beta}\right)}\Big ]_{\beta = -i\g n s^2 \lambda}.
\end{align}

Let's  first take the derivative, with respect to $\beta$, of the numerator and write part of the residue as
\begin{equation}\label{term1}
\text{term1}=-i\frac{e^{-\g ns^2\lambda}}{(\g ns^2)^{n-3}\prod_{j=1}^{n-3}(\lambda_j-\lambda)}.
\end{equation}
Next, differentiating the denominator, we obtain the other part of the residue as
\begin{align}\label{term2}
\text{term2}
&= e^{-i\beta }\Big [ 
\frac{i}{(\g n s^2 \lambda_1 - i\beta)^2 \prod_{j=2}^{n-3}\left({\g n s^2\lambda_j}-{i\beta}\right)} \nonumber\\
&+ \frac{i}{(\g n s^2 \lambda_2 - i\beta)^2 \prod_{j=1,j\neq 2}^{n-3}\left({\g n s^2\lambda_j}-{i\beta}\right)} + \ldots \Big ]\bigg|_{\beta = -i\g n s^2 \lambda} \nonumber\\
&=i \frac{e^{-\g n s^2 \lambda}}{(\g n s^2 )^{n-2}\prod_{j=1}^{n-3}(\lambda_j-\lambda)} \bigg[ \frac{1}{\lambda_1-\lambda}+\frac{1}{\lambda_2-\lambda}+\frac{1}{\lambda_3-\lambda}+\ldots \bigg].
\end{align}

A comparison of term1 and term2, given by Eqs. (\ref{term1}) and (\ref{term2}), respectively, shows that there is an extra factor of $s^2$ in the denominator of term2. Thus, term2 can be ignored compared to term1. Therefore, we obtain the distribution function for $s^2$ as
\begin{align}\label{RgAR}
P(s^2)&=\frac{|\mL_0| n^2 \g^2}{\prod_{j=1}^{n-3}(\lambda_j-\lambda)} s^3e^{-\g n \lambda s^2}\equiv Cs^3e^{-\tilde{\alpha} s^2}
\end{align}
where $C$ is the normalization constant that we will fix later, and $\tilde{\alpha}=\g n\lambda$. Note that the lowest eigenvalue for the $n$-dimensional Kirchoff's matrix is proportional to $1/n$, thus $n\lambda \sim \mathcal{O}(1)$. 

Now, $s^2=\L_1+\L_2$ and the aspect ratio $r=\sqrt{\L_1}/\sqrt{\L_2}$. Moreover, we have $\sqrt{\L_1}\sqrt{\L_2}=A$, where $A$ is the average area. Since the distribution of cell area is sharply peaked (Fig. 2e in the main text), as the cell division and apoptosis are slow processes, $A$ can be taken as a constant. Therefore, using the last two relations in the first, we obtain $s^2=A(r+1/r)$. Thus, we obtain from Eq. (\ref{RgAR}), the distribution of $r$ as
\begin{equation}
P(r)=\frac{1}{\mathcal{N}}\left(r+\frac{1}{r}\right)^{3/2}(1-\frac{1}{r^2})e^{-\alpha(r+\frac{1}{r})}
\end{equation}
where the condition that total probability must be unity sets the normalization constant $\mathcal{N}$ as
\begin{equation}
\mathcal{N}\simeq \frac{\Gamma(5/2)}{\alpha^{5/2}}-W(5/2)\,\, _1F_1(5/2,7/2,-2\alpha),
\end{equation}
where, $W(x)=2^x/x$ and $_1F_1(a,b,c)$ is the Kummer's confluent Hypergeometric function \cite{erdelyibook}.
Using the value of $\g$, we obtain $\alpha\propto \lambda_P(1-KP_0)/T$.
For the analysis of the theory, a symbolic software, such as ``Mathematica'' \cite{mathematica} is helpful. We note here that the relation between $sd$ and $\rav$ is quite complex, though for most practical purposes it can be taken as a straight line: $sd\simeq 0.71\, \rav-0.75$.

\subsection{Distribution for area}
\label{areaderivation}
Equation (\ref{RgAR}) gives the distribution of the radius of gyration of different cells in a monolayer. Using this equation, we now derive the distribution of area, $A$. Since, AR and $A$ can vary independently, we can use $s^2\propto A$, as we are interested in the functional form. Therefore, Eq. (\ref{RgAR}) gives
\begin{equation}\label{distA_noconstraint}
P(A)\sim A^{3/2}\exp[-\beta A]
\end{equation}
where $\beta$ is a constant related to $\tilde{\alpha}$. Note that we ignored the area term in the energy function, $\mathcal{H}$, in the derivation of Eq. (\ref{RgAR}). In a confluent cellular monolayer, the individual cell area must obey the strong geometric constraint of confluency; thus, it becomes crucial for the area distribution. However, mathematically imposing this constraint is a challenging geometric problem for random patterns, and no exact result exists yet. Weaire {\it et al} \cite{weaire1986} proposed a phenomenological implementation of this constraint as a polynomial in the area; this remains one of the simplest possible ways to date to deal with this constraint \cite{gezer2021}. Keeping only one term of this polynomial for simplicity, we can write this constraint as $f(A)\sim A^{\nu}$ \cite{weaire1986,gezer2021}. Using this in Eq. (\ref{distA_noconstraint}), we obtain
\begin{equation}\label{distA}
P(A)\sim A^{\mu-1}\exp[-\beta A],
\end{equation}
where we have defined $\mu=\nu+5/2$. From Eq. (\ref{distA}), we obtain the average area, $\bar{A}=\mu/\beta$. Thus, we obtain the normalized distribution for the scaled area, $a=A/\bar{A}$, as
\begin{equation}\label{distarea}
P(a)=\frac{\mu^\mu}{\Gamma(\mu)} a^{\mu-1}\exp[-\mu a],
\end{equation}
this is the well-known $k$-Gamma function, defined in Ref. \cite{aste2008}, and usually denoted with the variable $k$. This same function has been used in fitting the scaled AR, $r_s$, data in different existing experiments and simulations. Therefore, to avoid confusion with $k$, which is obtained fitting the $r_s$ data, we have used $\mu$ to define the distribution function for $a$. Since $\mu$ comes from the constraint of confluency, it should be independent of $\lambda_P$. Our simulation results within both the CPM and the VM (Fig. 2(f), in the main text) support this hypothesis.

\subsection{Simulation details}
\label{simdetails}

We have verified our theory via simulations of two distinct models: the CPM and the VM. In both the simulations, we have used the energy function $\mathcal{H}$, given in Eq. (1) in the main text.
The CPM is a lattice-based model. The underlying lattice has some effects on the quantitative aspects of the model; for example, on the square lattice, the object with the minimum possible perimeter for a given area is a square. However, the qualitative behaviors are independent of the lattice. To ascertain that the simulation results are independent of the lattice, we have simulated the CPM on two different lattices: the square lattice and the hexagonal lattice.

For the glassy dynamics, the geometric restriction leads to two different regimes, the low-$P_0$, and the large-$P_0$ regime. However, the large-$P_0$ regime characterizes quite large adhesion compared to the cortical contractility; this regime, we feel, is not relevant for the experiments. Therefore, we present most results when $P_0$ is not too large. 
Here, we briefly discuss the different models.

{\bf Cellular Potts Model (CPM):} 
In CPM, dynamics proceeds by stochastically updating one boundary lattice site at a time. In a Monte-Carlo simulation, we accept a move with a probability, $P(\sigma \rightarrow \sigma^\prime) = e^{-\frac{\Delta \mathcal{H}}{T}}$, where we have set Boltzmann constant $k_B$ to unity, $\Delta \mathcal{H}$ is the change in energy going from one configuration to the other. $\sigma$ and $\sigma^\prime$ are the cell indices of the current cell and the target cell indices, respectively. Unit of time refers to $M$ such elementary moves, where $M$ is the total number of lattice sites in the simulation. 

{\it \underline{With square lattice} :}
We have implemented CPM with square lattice in Fortran 90 and followed a \textsl{Connectivity Algorithm} developed in \cite{durand2016} to prevent fragmentation of cells \cite{souvik2021}. We have chosen a simulation box of size $120\times120$ with $360$ total cells in the system. The average area of cells in the system is $40$, and the minimum possible perimeter on a square lattice with this area is $26$. Unless otherwise specified, we always start with an initial configuration where each cell is rectangular with a size $5\times 8$. We first equilibrate the system for $8\times 10^5$ MC time steps before collecting the data of aspect ratio. The variables that characterize the system are $\lambda_A$, $\lambda_P$, $P_0$, and $T$.
We have ignored cell division and apoptosis, as discussed in the main text. However, to test the effects of these processes, we have included them within this model and present a specific set of results in Sec. \ref{divapop}. 

To obtain the relaxation time, $\tau$, we have calculated the overlap function, $Q(t)$, defined as
\begin{equation}
Q(t)=\overline{\frac{1}{N}\sum_{\sigma=1}^N \langle W(a-|\mathbf{X}_{cm}^\sigma(t+t_0)-\mathbf{X}_{cm}^\sigma(t_0)|)\rangle_{t_0}},
\end{equation}
where $X_{cm}^\sigma(t)$ is the center of mass at time $t$ of a cell with index $\sigma$, $W(x)$ is a Heaviside step function
\begin{align}
W(x)=\begin{cases}
1  & \text{if } x\geq 0\\
0 & \text{if } x<0,
\end{cases}
\end{align}
$\langle \ldots\rangle_{t_0}$ denotes averaging over initial times and the overline implies ensemble average. The parameter $a$ is related to the vibrational motion of the particles (here cells) inside the cage formed by its neighbors. We have set $a=1.12$ \cite{souvik2021}.  We have taken 50 $t_0$ averaging and 20 configurations for ensemble averaging.

{\it \underline{With Hexagonal lattice} :}
To simulate CPM on the hexagonal lattice, we have used an open-source application CompuCell3D \cite{swat2012,zajac2003}.  CPM is the core of CompuCell3D. We have simulated the 2D confluent cell monolayer with periodic boundary conditions using the same energy function, Eq. (1), in the main text. In this simulation, we have taken $128\times128$ simulation box size with 361 cells. Initially, Cells are of three different types of areas and perimeters. We have started with a rectangular slab of the cell monolayer. We first equilibrate the system for $10^5$ MC time steps before collecting the data. Fragmentation is allowed in these simulations; however, we have restricted ourselves in the low $T$ regime, where fewer cells are fragmented. While calculating the PDFs, we have eliminated the fragmented cells.

{\bf Vertex Model (VM):}
Vertex models(VM) can be viewed as the continuum version of the CPM. Within VM, vertices of the polygons are the degrees of freedom, and cell edges are defined as straight lines (or lines with a constant curvature) connecting  between vertices \cite{farhadifar2007,fletcher2014}. Within the Monte-Carlo simulation \cite{wolff2019}, dynamics proceeds by stochastically updating each vertex position of all the cells by a small amount $\delta r$. We accept the move with a probability $P(\mathcal{C} \rightarrow \mathcal{C}^\prime) = e^{-\frac{\Delta \mathcal{H}}{T}}$ where we have set Boltzmann constant $k_B$ to unity, $\Delta \mathcal{H}$ is the change in energy going from $\mathcal{C}\rightarrow\mathcal{C}^\prime$. Initially, we start will 1024 cells having equal area $(=1.0)$ and perimeter $(=3.72)$. We have equilibrated the system for $2\times 10^6$ MC steps before the start of collecting data.

\begin{figure}[h!]
	\label{ARSARhexSquareSM}
	\includegraphics[width=16cm]{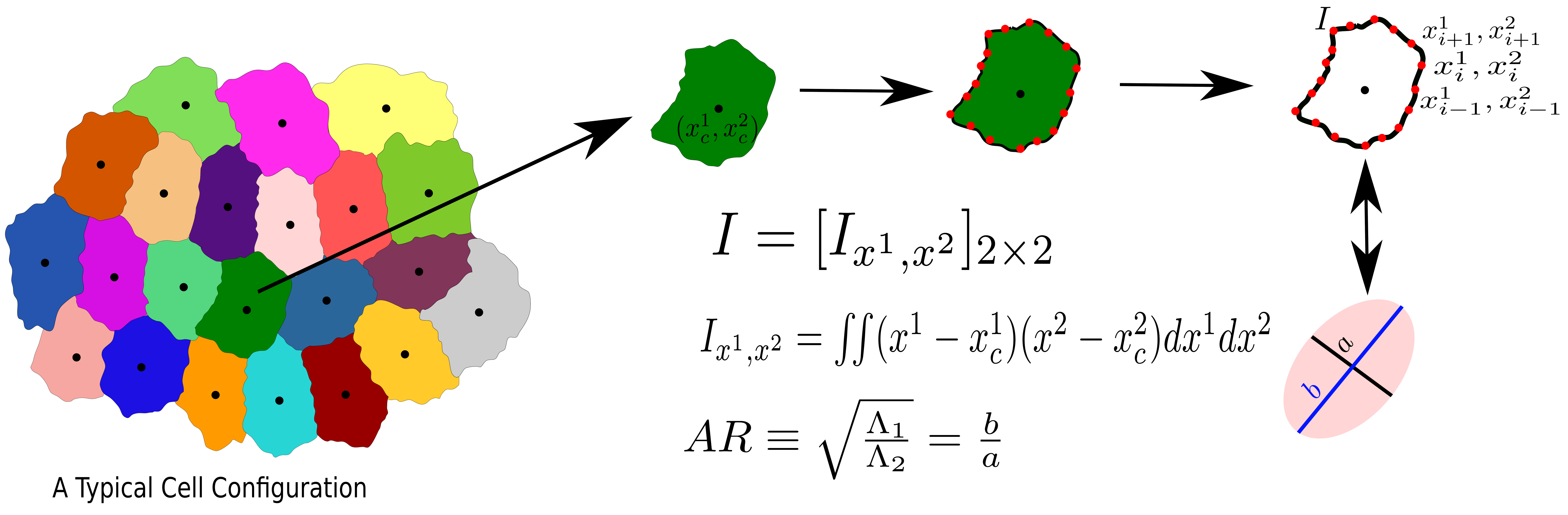}
	\caption{The process to calculate the AR of a Cell. We first obtain the moment of inertia, $\mathbf{I}$, in a frame of reference whose origin coincides with the center of mass of the cell, then diagonalize $\mathbf{I}$, and calculate the AR, $r$, as the square root of the ratio of the two eigenvalues. }
	\label{ARprocessSM}
\end{figure}

\subsection{Procedure to calculate AR in our simulation}
\label{ARprocedure}
As stated in the main text, we characterize the cell shape via AR. To calculate the AR, we follow Ref. \cite{atia2018} and briefly discuss the method here. Consider one particular cell, represented by the set of points $\{x_i^1,x_i^2\}$, as schematically shown in Fig. \ref{ARprocessSM}. We first calculate the moment of inertia in a frame of reference whose origin coincides with the center of mass $(x_c^1,x_c^2)$ of the cell. The moment of inertia, $\mathbf{I}$, is a $2\times2$ tensor in spatial dimension two. We next diagonalize $\mathbf{I}$ and obtain the two eigenvalues, $\L_1$ and $\L_2$. Without loss of generality, we assume that $\L_1\geq \L_2$, and obtain the AR, $r=\sqrt{\L_1/\L_2}$. This process can be viewed as approximating the cell with an ellipse with the major and minor axes given by $\sqrt{\L_1}$ and $\sqrt{\L_2}$, respectively. For the CPM on the square lattice and the VM, we have written our own codes; for the \textsl{CompuCell3D} \cite{swat2012,zajac2003}, we have used \texttt{<Plugin Name="MomentOfInertia"/>} in the .XML file and using the example of \texttt{Demos/MomentOfInertia}, we have calculated the lengths of semiaxes in the \texttt{Steppables.py} file.  We have obtained the AR of a cell by taking the ratio of lengths of semi-major and semi-minor axes.

\begin{figure}
	\includegraphics[width=12cm]{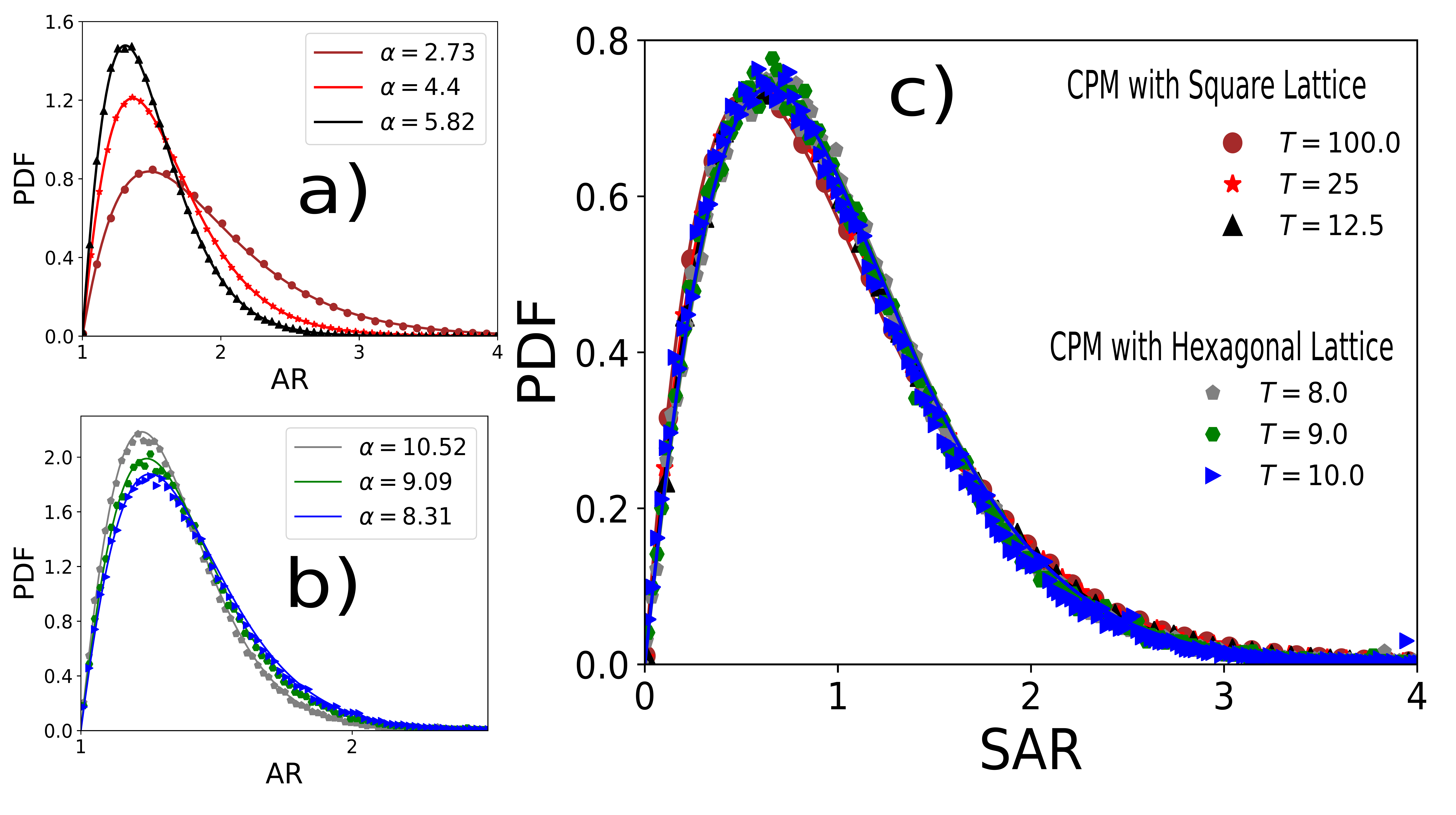}
	\caption{Comparison among the CPM simulations on a square and hexagonal lattice. We have used  $\lambda_A = 1.0$, $\lambda_P=0.5$, $T$ values are quoted in (c). (a) PDF of $r$ on square lattice with $P_0=26$, (b) PDF of $r$ on hexagonal lattice with $P_0=25$. In both (a) and (b), symbols are simulation data, and lines represent the fits with Eq. (6) in the main text. Values of $\alpha$ are given in the figures. (c) PDFs for the $r_s$, for the same data as in (a) and (b), show a virtually universal behavior. }
	\label{ARSARhexSquareSM}
\end{figure}

\subsection{Distribution of AR is independent of lattice type}
\label{hexresults}
In the main text, we have presented the results for the CPM on a square lattice and the VM, the latter being a continuum model. The agreements between the two models show that the results are independent of the lattice. As further evidence of this lattice independence, we present the results within the CPM on the square and the hexagonal lattices in Fig. \ref{ARSARhexSquareSM}. We show that our analytical theory agrees with simulations on both lattices.

Figures \ref{ARSARhexSquareSM}(a) and (b) show representative plots comparing the PDFs of $r$ within the CPM with the square and the hexagonal lattices, respectively. As discussed in the main text, our theory predicts almost universal behavior for the PDFs of $r_s$ (Fig. \ref{ARSARhexSquareSM} c). This result shows that qualitative behaviors within the CPM are lattice-independent and agree with our analytical theory.

\subsection{Distribution of AR does not depend on $\lambda_A$}
\label{lamAAR}
\begin{figure}[h!]
	\includegraphics[width=12cm]{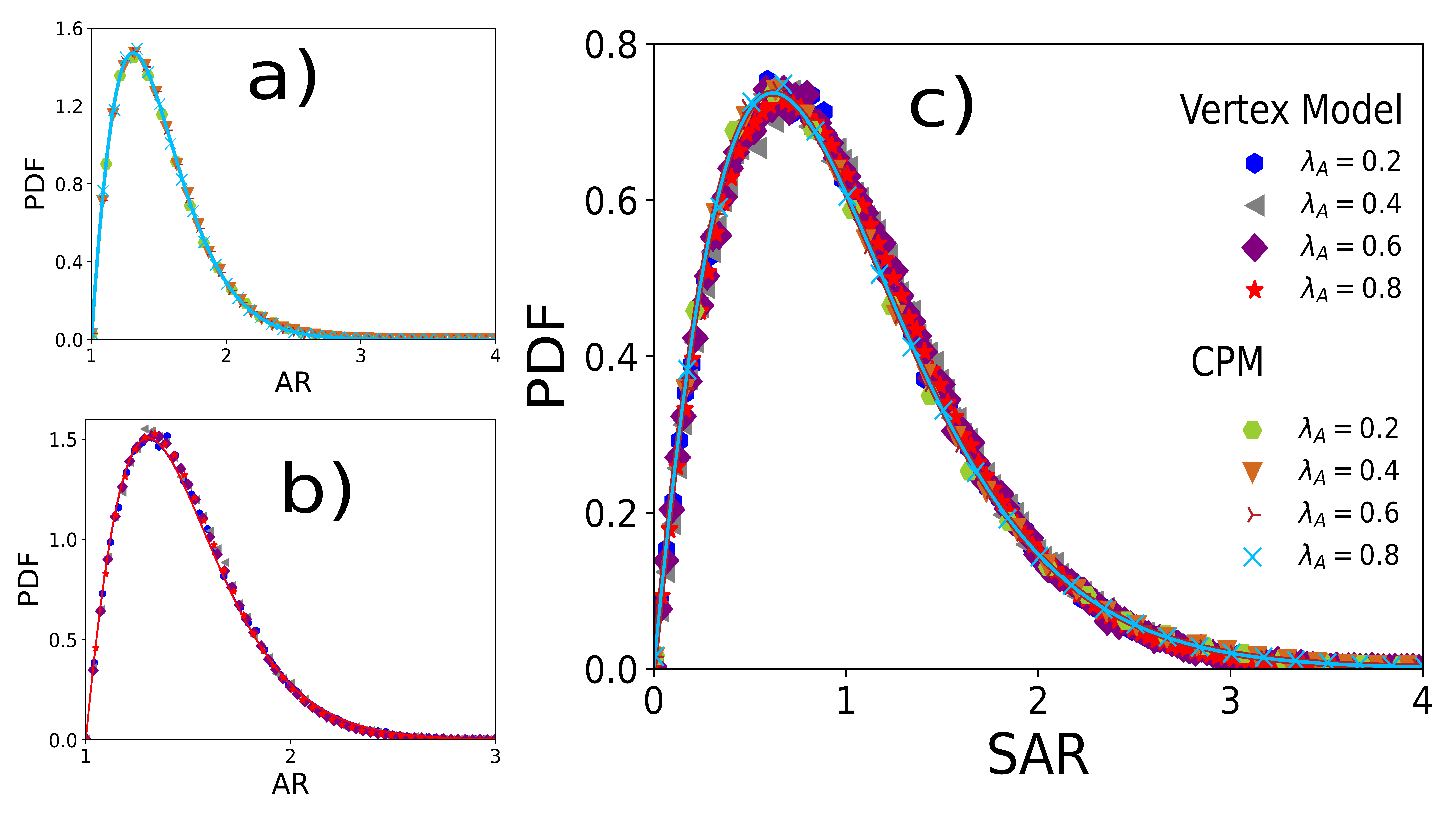}
	\caption{PDFs of $r$ at different values of $\lambda_A$ within the CPM and the VM. Values of $\lambda_A$ are quoted in (c). (a) PDF of AR within the CPM for $\lambda_P=0.5$ and $T=12.5$ and (b) within the VM for $\lambda_P=0.02$ and $T=0.0025$ (c) PDFs of $r_s$, for the same data as in (a) and (b), show a virtually universal behavior. Symbols are simulation data, and lines represent the fits with Eq. (6) with $\alpha$ given in the main text in inset of Figs. 2(a) and (b).}
	\label{ARSARdifflamASM}
\end{figure}

One of the arguments within our mean-field theory is the following: since AR can be independent of $A$ and the shape should have the dominant contribution from the cortical properties, we can assume that individual cells satisfy the area constraint and ignore the area term in $\mathcal{H}$, Eq. (1) in the main text. We have shown that the area distribution is sharply peaked around the average area (Fig. 2(e) in the main text, and Fig. \ref{areaPDFVM}). We have also shown in the insets of Figs. 2(a) and (b) that $\alpha$ does not depend on $\lambda_A$, both within the CPM and the VM. 
We show in Fig. \ref{ARSARdifflamASM} that the PDFs of $r$ are almost overlapping with each other at different values of $\lambda_A$. Figure \ref{ARSARdifflamASM}(c) shows the PDFs for $r_s$.

\subsection{Distribution of AR including cell division and apoptosis in the CPM simulation}
\label{divapop}
\begin{figure}
	\includegraphics[width=16cm]{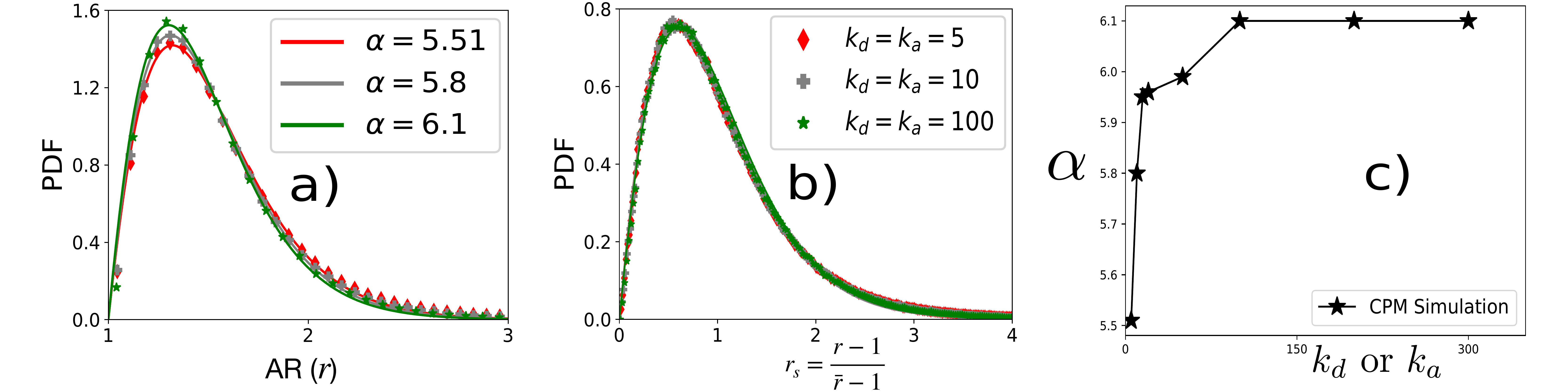}
	\caption{PDFs of AR in the presence of cell division and apoptosis. (a) PDFs of AR at different values of $k_d = k_a$ (quoted in b)  within CPM for $\lambda_P=0.5$ and $T=12.5$. (b) PDF of $r_s$ corresponding to the same data as in (a). (c) $\alpha$ vs $k_d$ or $k_a$  quickly reaches a constant with increasing $k_a$.}
	\label{ARSARwithcelldivapopSM}
\end{figure}

We have ignored cell division and apoptosis in our theory since the rates of these processes are extremely low. However, even when these processes are significant, it is easy to see that their effects can only enter via $\alpha$. The reason behind this is that the other, algebraic part, comes from the topology of the cell perimeter, that it is a closed-looped object; this property must remain the same even in the presence of these processes. Therefore, all the conclusions of our theory should remain valid even when the rates of division and apoptosis are significant.

To test this result, we have included cell division and apoptosis in our simulations. We keep the rate of division, $k_d^{-1}$, and apoptosis, $k_a^{-1}$, the same such that the average number of cells remains the same. Every $k_d$ time step, we randomly select a cell and divide it into two, with a randomly chosen division plane. To decide the division plane, we first calculate the center of mass (CoM) and then chose a point on the perimeter in a random direction. The line connecting the CoM and this point gives the division plane. The area of these two daughter cells then grows till it becomes of the order of $A_0$ (Eq. (1) in the main text). To avoid diving a cell that has just undergone division, we impose a cut-off on area ($\geq 32$) for selecting a cell for division. For apoptosis, every $k_a$ time step, we randomly choose a cell and assign it a target area $A_0=0$. We also relax the constraint that does not allow fragmentation in our simulation for the cells undergoing apoptosis.

We show the PDF of $r$ at three representative values of $k_a$ in Fig. \ref{ARSARwithcelldivapopSM}(a), and the corresponding PDFs for $r_s$ are shown in Fig. \ref{ARSARwithcelldivapopSM}(b). $k_a$ and $k_d$ are presented in units of the inverse of the Monte-Carlo time. We have fitted the PDFs of $r$ with Eq. (6) of the main text, and the values of $\alpha$ at different $k_a$ are shown in Fig. \ref{ARSARwithcelldivapopSM}(c). We emphasize two observations: first, the PDFs of $r$ indeed fit well with Eq. (6) of the main text, and second, that $\alpha$ attains a constant value quite fast with increasing $k_a$, that is, decreasing the rate of division or apoptosis.

\begin{figure}
	\includegraphics[width=14cm]{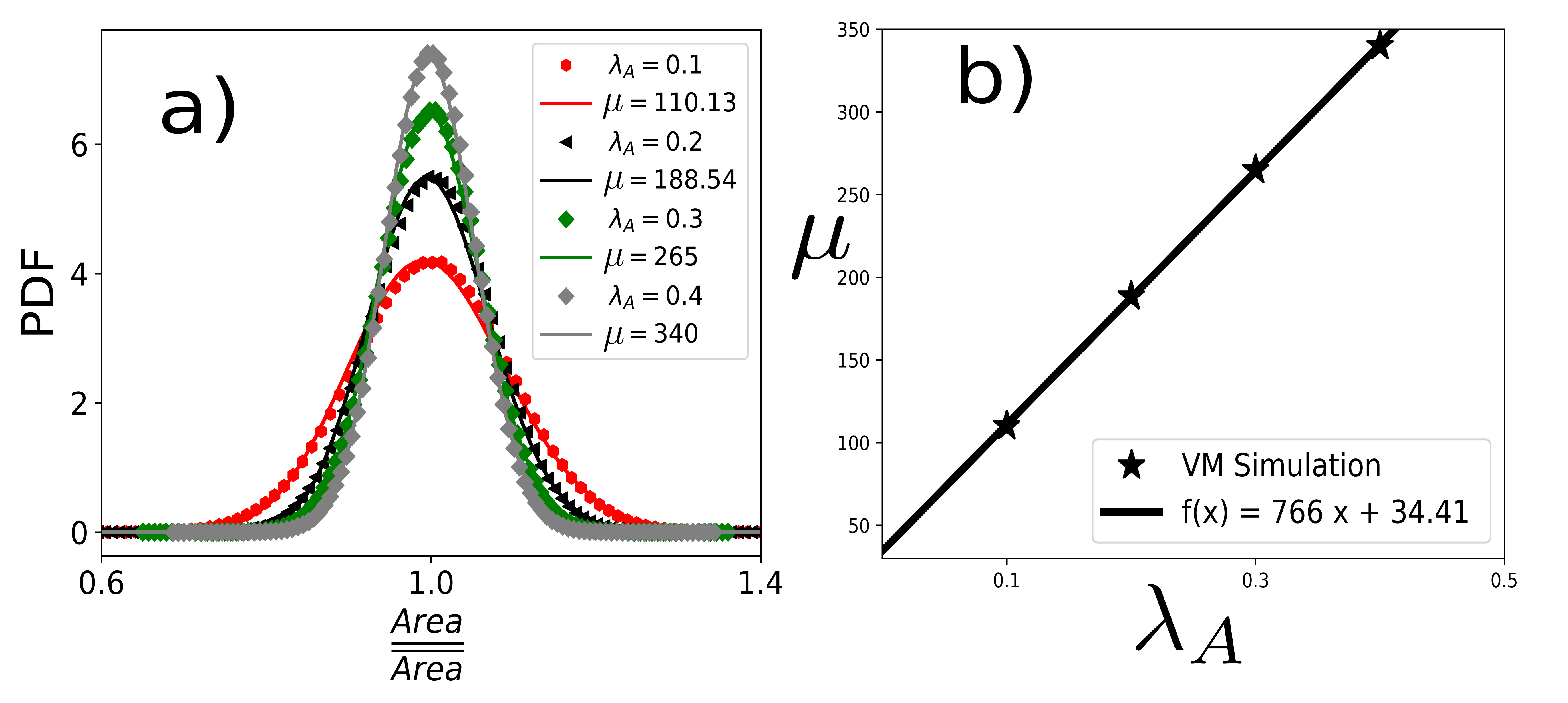}
	\caption{Scaled area distribution within the VM. (a) PDF of the scaled area, $a$, within the VM at different $\lambda_A$ and fixed $P_0 = 3.7$, $\lambda_P = 0.02$, and $T = 0.0025$. Symbols are simulation data, and lines represent the fits with Eq. (7) of the main text. (b) $\mu$ linearly increases with $\lambda_A$.}
	\label{areaPDFVM}
\end{figure}
\subsection{Scaled area distribution within the VM with varying $\lambda_A$}
\label{areadistVM}

In the main text, we have shown the distribution of scaled area, $a$, for different values of $\lambda_A$ within CPM and their comparison with the analytical theory, Eq. (7). Here, in Fig. \ref{areaPDFVM}(a), we show the same distribution within the VM simulation and comparison with our theory. 
Figure \ref{areaPDFVM}(b) shows that $\mu$ linearly increases as $\lambda_A$ increases, similar to the result within the CPM (Fig. 2e, in the main text).

\begin{figure}
\includegraphics[width=15cm]{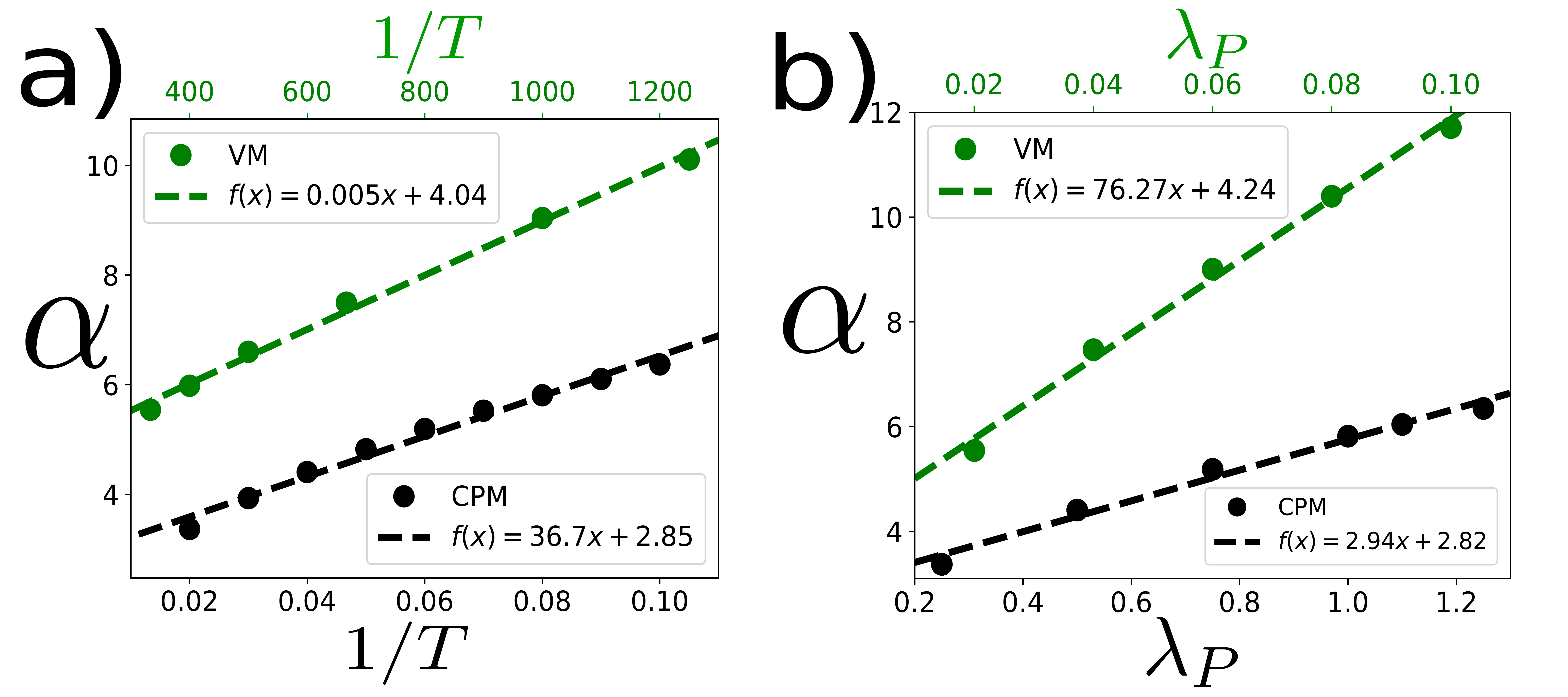}
\caption{Behavior of $\alpha$ with $T$ and $\lambda_P$. (a) $\alpha$ increases almost linearly as a function of $1/T$. Parameters for the simulations with the VM (represented by the green symbols and line) $\lambda_A =0.5$, $\lambda_P =0.02$, $P_0 =3.7$. Within the CPM (black): $\lambda_A =1.0$, $\lambda_P =0.5$, and $P_0 =26$. (b) $\alpha$ increases almost linearly as a function of $\lambda_P$.  For the VM (green): $\lambda_A =0.5$, $T =0.003$, $P_0 =3.7$. For the CPM (black): $\lambda_A =1.0$, $T =25$ ,$P_0 =26$. Points represent simulation data and dashed lines represent the fits.}
\label{alphalamPT}
\end{figure}

\begin{figure}[h]
	\includegraphics[width=10cm]{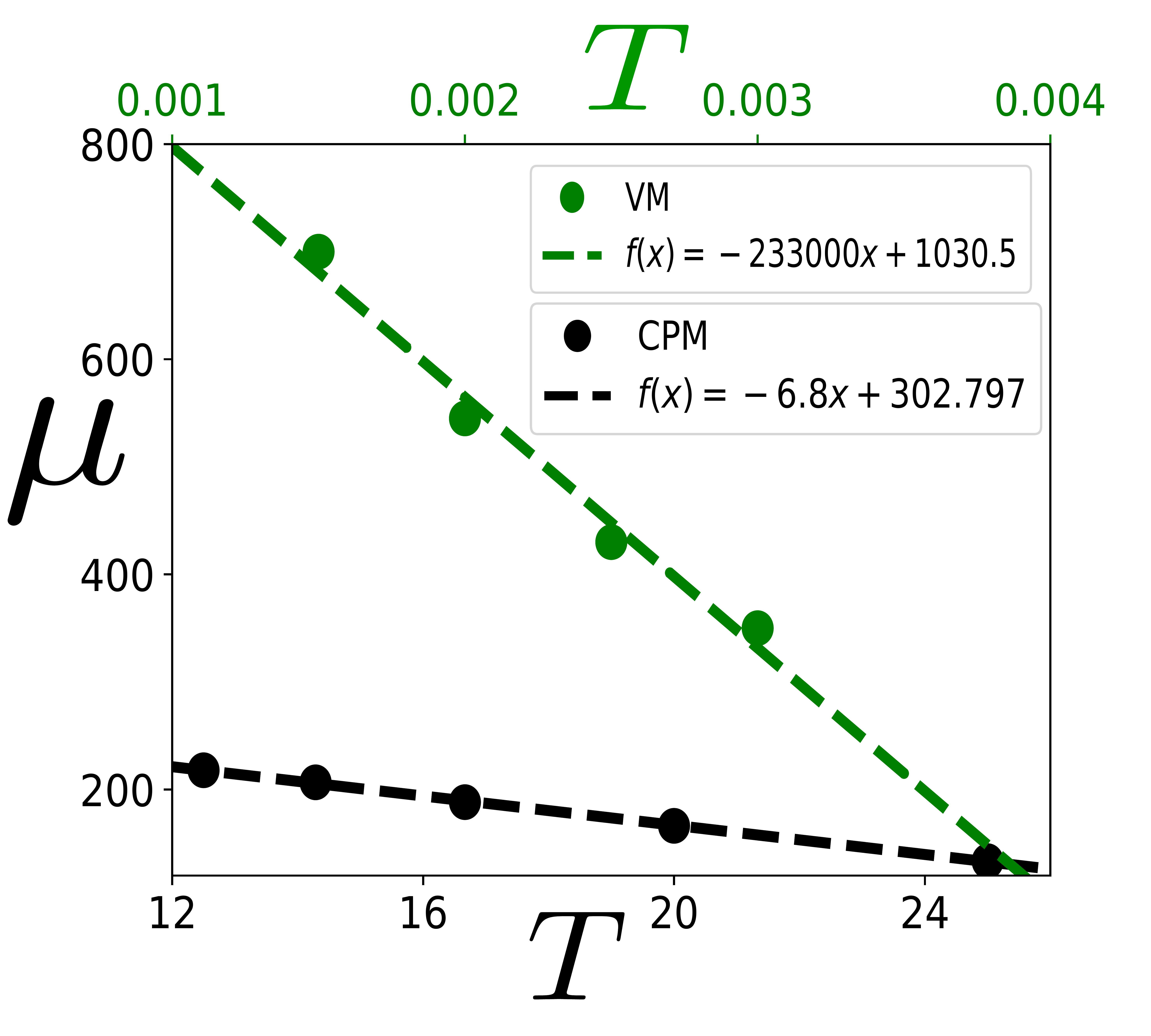}
	\caption{The behavior of $\mu$ as a function of $T$. The coefficient $\mu$, in Eq. (7) in the main text, characterizes the PDF of scaled area, $a$. We show that $\mu$ decreases almost linearly as $T$ increases within both the CPM and the VM. In these simulations, the values of the parameters are as follows. For VM: $P_0=3.7$, $\lambda_A=0.5$, and $\lambda_P=0.02$; for CPM: $P_0=26$, $\lambda_A=1.0$, and $\lambda_P=0.5$.}
	\label{muvsT}
\end{figure}
\subsection{Dependence of $\alpha$ on $\lambda_P$ and $T$ within the CPM and the VM}
\label{alpha_individual}
In the main text, we have shown the linear dependence of $\alpha$ on $\lambda_P/T$: within both the CPM and the VM (Fig. 2d in the main text). Here, we show how $\alpha$ varies within both models when we change $\lambda_P$ and $T$, individually. Figure \ref{alphalamPT}(a) shows the behavior of $\alpha$ as a function of $1/T$ and Fig. \ref{alphalamPT}(b) as a function of $\lambda_P$: $\alpha$ increases almost linearly as $1/T$ or $\lambda_P$ increases, in agreement with our theory.

\subsection{Dependence of $\mu$ on $T$}
\label{muonT}
Figure 2(f) in the main text shows that $\mu$ remains almost constant with varying $\lambda_P$. This behavior is expected since $\mu$ comes from the constraint of confluency and should be related to the area, whereas $\lambda_P$ is related to the perimeter constraint. On the other hand, from the same argument, we also expect $\mu$ to depend on $\lambda_A$; as shown in Fig. 2(e) in the main text, it increases linearly with $\lambda_A$. In addition, since $T$ controls the fluctuation in the system, we also expect $\mu$ to change with $T$; this dependence leads to different distributions of $a$ in epithelial systems. We show in Fig. \ref{muvsT} that $\mu$ almost linearly decreases as $T$ increases.

\begin{figure}
	\includegraphics[width=12cm]{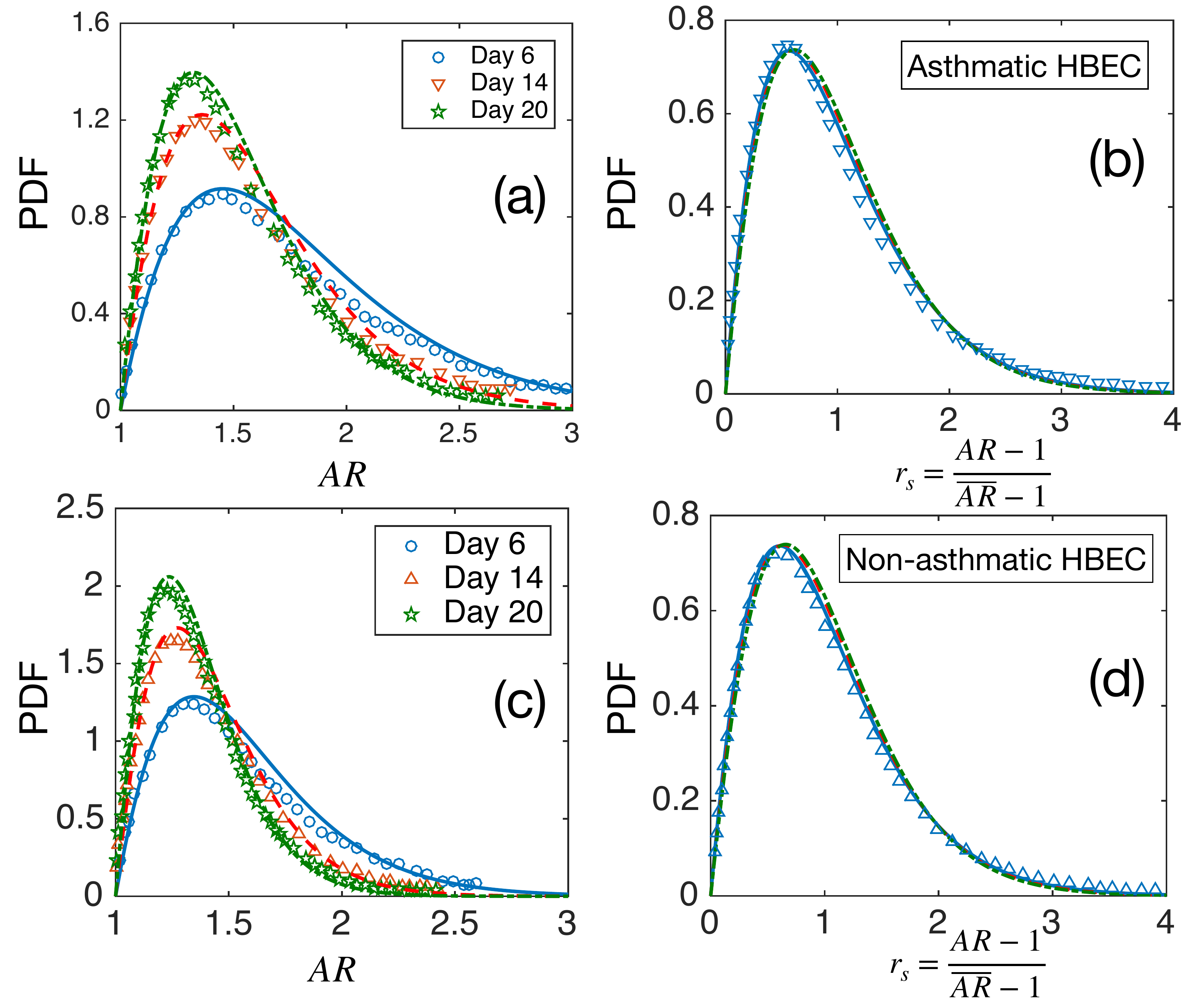}
	\caption{Comparison of theory with experiments on HBEC cell monolayer. (a) Probability distribution function (PDF) of AR for asthmatic HBEC cells. Symbols are data taken from Ref. \cite{atia2018}, and the lines represent fits with our theory, Eq. (6) in the main text. Values of $\alpha$ are shown in the main text, Table I. (b) PDF of the scaled AR, $r_s$ for the asthmatic HBEC cells. Lines are the theoretical predictions using the values of $\alpha$ obtained from the fits in (a), symbols represent one set of experimental data \cite{atia2018}. (c) and (d) The same as in (a) and (b), but for non-asthmatic HBEC cells. }
	\label{hbecdata}
\end{figure}

\subsection{Fits with the HBEC data of AR}
\label{hbecdatafit}

We have compared our theory for the AR, $r$, and the scaled AR, $r_s$, with the experimental data for three different systems: MDCK cells, asthmatic HBEC, and non-asthmatic HBEC cells. These data are taken from Ref. \cite{atia2018}, where PDFs of $r$ and $r_s$ were presented as a function of maturation. As discussed in the main text, we have selected the data at three different times, and the comparison for the MDCK cells are shown in Fig. 3(a) in the main text. Here, in Fig. \ref{hbecdata}, we present the comparisons for the asthmatic and non-asthmatic HBEC cells. We have collected only one set of experimental data for the PDF of $r_s$ in both the asthmatic and non-asthmatic systems, as all the data nearly overlap.
The experimental data from different papers are collected using the software ``WebPlotDigitizer" \cite{ankit2020} that allows reading off the data from a digital plot.

\bibliography{SupMat_aspectratioref.bib}